\documentclass[aps,twocolumn,longbibliography,superscriptaddress]{revtex4-1}
\pdfoutput=1

\usepackage{amsmath,amssymb,mathtools,bm,color,graphicx,empheq}

\newcommand{\real}{\operatorname{Re}}
\newcommand{\imag}{\operatorname{Im}}
\newcommand{\sinc}{\operatorname{sinc}}

\begin{document}

%\title{Quantum backflow current in a ring:\\ optimal bounds, fractal dimension, analytical examples}
\title{Quantum backflow current in a ring: Optimal bounds and fractality}

\author{Arseni Goussev}
\affiliation{Section of Mathematics, University of Geneva, Rue du Conseil-G\'en\'eral 7-9, 1205 Gen\`eve, Switzerland}
\affiliation{Quantum Physics Corner Ltd, 20--22 Wenlock Road, London N1 7GU, United Kingdom}

\author{Felix Quinque}
\author{Jaewoo Joo}
\author{Andrew Burbanks}
\affiliation{School of Mathematics and Physics, University of Portsmouth, Portsmouth PO1 3HF, United Kingdom}

\date{\today}

\begin{abstract}
	The probability density of a quantum particle moving freely within a circular ring can exhibit local flow patterns inconsistent with its angular momentum, a phenomenon known as quantum backflow. In this study, we examine a quantum particle confined to a ring and prepared in a state composed of a fixed (yet arbitrary) number of lowest energy eigenstates with non-negative angular momentum. We investigate the time-dependent behavior of the probability current at a specified point along the ring's circumference. We establish precise lower and upper bounds for this probability current, thereby delineating the exact scope of the quantum backflow effect. We also present an analytical expression for a quantum state that yields a record-high backflow probability transfer, reaching over $95\%$ of the theoretical bound. Furthermore, our investigation yields compelling numerical and analytical evidence supporting the conjecture that the current-versus-time function associated with states maximizing backflow probability transfer forms a fractal curve with a dimension of $7/4$. The observed fractality may provide a characteristic, experimentally-relevant signature of quantum backflow near the probability-transfer bound.
\end{abstract}

\maketitle

\section{Introduction}

In classical mechanics, a particle always moves in the direction of its momentum, which serves as a measure of the intensity of the particle's motion. In quantum mechanics, however, the situation can be strikingly different: the probability density of a quantum particle may in fact flow against its momentum. This intriguing phenomenon, initially recognized within the context of the arrival time problem \cite{All69time-c, Kij74time, Wer88Wigner}, is known as quantum backflow (QB).

The first systematic examination of QB was conducted by Bracken and Melloy \cite{BM94Probability}. They examined the time-evolution of the probability density of a free particle on a line constrained to move with positive momentum and addressed the (classically impossible) flow of the probability density in the negative direction. Their most notable finding was the fact that the total probability transported through a fixed spatial point cannot exceed a certain threshold, commonly known as the Bracken-Melloy (BM) bound. The BM bound is independent of the particle's mass, the observation time interval, or the Planck constant. Numerical estimates indicate the BM bound to approximately equal 0.0384517 \cite{EFV05Quantum, PGKW06new}. While the exact value of the BM bound remains unknown, it has been recently proven to lie  between 0.0315 and 0.0725~\cite{TLN23Quantum}.

The phenomenon of QB has been addressed in the literature across various scenarios and formulations. Among the problems explored are QB against a constant force \cite{MB98velocity}, the spatial extent of the backflow proability current \cite{EFV05Quantum, Ber10Quantum, BCL17Quantum}, the relationship between QB and the arrival time problem \cite{MPL99Arrival, ML00Arrival, DEHM02Measurement, HBLO19Quasiprobability}, QB for rotational motion \cite{Str12Large, PPR20Angular, Gou21Quantum, BGS23Unbounded}, QB in many-particle systems \cite{Bar20Quantum, MM20Quantum} and in the presence of noisy and dissipative environments \cite{AGP16Quantum, MM20Dissipative, MM20Erratum}, backflow in relativistic systems \cite{MB98Probability, SC18Quantum, ALS19Relativistic, BBA22Backflow, BPPR23Quantum}, QB across a black hole horizon \cite{BG21Quantum}, and the classical limit of QB \cite{YHHW12Analytical, BM--Remarks, Bra21Probability, MM22Different}. Multiple analytical examples of states exhibiting probability backflow have been constructed \cite{BM94Probability, EFV05Quantum, YHHW12Analytical, HGL+13Quantum, Chr24Design}. QB has been explored both in phase space \cite{BM94Probability, AGP16Quantum, MYDP21Experiment, BG21experiment} and as variations of the quantum reentry problem \cite{Gou19Equivalence, DT19Decay, Gou20Probability, Str24Quantum}. The reader is directed to Ref.~\cite{YH13introduction} for an elementary introduction to the phenomenon of QB and to Refs.~\cite{BM14Waiting, BFG+23Comment, BM23Comment} for non-technical discussions of its physical interpretation and nonclassical character.

So far, direct experimental observation of QB remains elusive, despite existing proposals exploring methods to detect the effect using Bose-Einstein condensates \cite{PTMM13Detecting, MPM+14Interference}. The challenges associated with the experimental realization of QB for a particle on a line stem from two main factors. Firstly, only a minute portion of the overall probability, given by the BM bound, can theoretically be transported in the ``wrong'' direction. Secondly, the quantum states that exhibit backflow probability transfer near the BM bound are characterized by their infinite spatial extent and infinite energy \cite{YHHW12Analytical}, rendering them challenging to produce in a laboratory environment. However, although the first experimental demonstration of QB is still forthcoming, the effect has already been successfully simulated in classical optics experiments \cite{EZB20Observation, DGGL22Demonstrating, GDGL23Azimuthal} and on a genuine quantum computer~\cite{GJ24Simulating}.

A promising avenue for future experimental realization of QB involves a quantum particle rotating freely in a circular ring \cite{Gou21Quantum}. In this scenario, the particle is prepared in a superposition of energy and angular momentum eigenstates with non-negative angular momentum, and the observed quantity is the probability current through a fixed point on the ring. Contrary to classical mechanical predictions, the quantum mechanical probability current can assume negative values, thereby manifesting the phenomenon of QB. For this scenario, it has been demonstrated \cite{Gou21Quantum} that the total backflow probability transfer over a finite time interval can exceed the BM bound by more than threefold, reaching approximately 0.116816. Moreover, the particle-in-a-ring states exhibiting significant backflow probability transfer have been shown to possess finite energy (and, due to the nature of the problem, finite spatial extent), rendering them more suitable for experimental realization.

In this paper, we explore the QB phenomenon for a particle confined to a circular ring, scrutinizing the probability current at a fixed point along the ring's circumference. Our study yields two primary findings. Firstly, we establish optimal lower and upper bounds on the probability current for the most general state of the system, encompassing a fixed (yet arbitrary) number of lowest energy eigenstates with non-negative momentum. In other words, we determine the minimally and maximally attainable values of the probability current associated with the most general superposition of non-negative angular momentum states, all with energies not exceeding a specified (yet arbitrary) threshold. Secondly, we propose a conjecture regarding the time-dependence of the probability current of the state maximizing backflow probability transfer, or states characterized by backflow probability transfers nearing the theoretical limit. More specifically, we conjecture that this probability current is a fractal function of time, with a fractal dimension of $7/4$, and present compelling numerical and analytical evidence in support of our conjecture. We suggest the possibility that fractal characteristics may provide an experimental signature of QB near the probability-transfer bound, important for future investigations.

The paper is organized as follows. In Sec.~\ref{sec:prob_current}, we define the system and introduce the dimensionless probability current, which will serve as the central object of study throughout the rest of the paper. In Sec.~\ref{sec:optimal_bounds}, we derive the optimal bounds for the probability current. Section~\ref{sec:prob_transfer} is dedicated to summarizing crucial facts about backflow probability transfer, laying the groundwork for the subsequent discussion. Section~\ref{sec:bm_state} explores the time-dependence of the probability current associated with the backflow-maximizing state and presents the numerical calculation of its fractal dimension. In Sec.~\ref{sec:analytical_approximation}, we construct an accurate analytical approximation for the backflow-maximizing state and determine that the fractal dimension of its corresponding current-versus-time function is $7/4$, in good agreement with the numerical value obtained in Sec.~\ref{sec:bm_state}. (In brief, Sections~\ref{sec:prob_current} and \ref{sec:prob_transfer} primarily introduce the system and review relevant existing results, whereas Sections~\ref{sec:optimal_bounds}, \ref{sec:bm_state}, and \ref{sec:analytical_approximation} present original findings.) We summarize our findings and provide closing remarks in Sec.~\ref{sec:conclusion}.

\section{Probability current}
\label{sec:prob_current}

We consider a particle of mass $M$ freely moving on a circular ring of radius $R$. The wave function of the particle is denoted by $\psi(\theta, T)$ and satisfies the Schr\"odinger equation
\begin{equation}
	i \hbar \frac{\partial \psi}{\partial T} = -\frac{\hbar^2}{2 M R^2} \frac{\partial^2 \psi}{\partial \theta^2} \,.
\label{TDSE}
\end{equation}
Here, $\theta$ is the angle coordinate of the particle, and $T$ is time. The wave function is subject to periodic boundary conditions, $\psi(-\pi, T) = \psi(\pi, T)$. The general solution to Eq.~\eqref{TDSE} has the form of a linear combination of energy eigenstates
\begin{equation*}
	\phi_m(\theta, T) = \frac{1}{\sqrt{2 \pi}} e^{i m \theta - i E_m T / \hbar} \,,
\end{equation*}
each labelled by an integer $m \in \mathbb{Z}$ and characterized by the energy value
\begin{equation*}
	E_m = \frac{\hbar^2 m^2}{2 M R^2} \,.
\end{equation*}
Since $(-i \hbar \partial / \partial \theta) \phi_m = (m \hbar) \phi_m$, the eigenstates $\phi_m$ are also characterized by definite values of angular momentum, $m \hbar$.

In what follows, we only consider quantum wave packets with non-negative angular momentum. Thus, let $\psi_N$ be a wave packet comprised of eigenstates $\phi_m$ with $0 \le m \le N$, namely
\begin{equation}
	\psi_N(\theta, T) = \frac{1}{\sqrt{2 \pi}} \sum_{m=0}^N c_m e^{i m \theta - i E_m T / \hbar} \,.
\label{psi_N}
\end{equation}
The wave packet is assumed to be normalized to unity, $\int_{-\pi}^{\pi} d\theta \, |\psi_N(\theta, T)|^2 = 1$, implying that the expansion coefficients satisfy
\begin{equation}
	 \sum_{m=0}^N |c_m|^2 = 1 \,.
\label{norm}
\end{equation}
By construction, any angular momentum measurement conducted on $\psi_N$ is certain to yield a non-negative result.

The main focus of present study is the probability current $J_N$ through an arbitrarily fixed point on the ring, taken to be $\theta = 0$ for concreteness:
\begin{equation}
	J_N(T) = \left. \frac{\hbar}{M R^2} \imag \left\{ \psi_N^*(\theta, T) \frac{\partial \psi_N(\theta, T)}{\partial \theta} \right\} \right|_{\theta = 0} \,.
\label{J_N}
\end{equation}
QB occurs when the probability current is negative, i.e., when $J_N(T) < 0$ for some $T$.

It is convenient to introduce dimensionless time $t$ and dimensionless probability current $j_N$ as
\begin{equation}
	T = \frac{2 M R^2}{\hbar} t \,, \quad J_N(T) = \left( \frac{2 M R^2}{\hbar} \right)^{-1} j_N(t) \,.
\label{dimensionless}
\end{equation}
Substituting Eq.~\eqref{psi_N} into Eq.~\eqref{J_N}, and taking into account transformations~\eqref{dimensionless}, we obtain the following expression for the dimensionless probability current:
\begin{equation}
	j_N(t) = \frac{1}{2 \pi} \sum_{m,n = 0}^{N} c_m^* c_n (m + n) e^{i (m^2 - n^2) t} \,.
\label{j_N}
\end{equation}

\section{Optimal bounds on the probability current}
\label{sec:optimal_bounds}

How small and how large can the probability current possibly be? In this section we answer this question by showing that
\begin{widetext}
\begin{equation}
	\frac{N (N + 1)}{4 \pi} \left( 1 - \sqrt{\frac{4 N + 2}{3 N}} \right) \le j_N(t) \le \frac{N (N + 1)}{4 \pi} \left( 1 + \sqrt{\frac{4 N + 2}{3 N}} \right) \,,
\label{key_result_1}
\end{equation}
\end{widetext}
for all $t$, and by finding the expansion coefficients $\{ c_m \}_{m=0}^N$ of the states corresponding to the extreme values of the current.

We begin by rewriting Eq.~\eqref{j_N} as an expectation value:
\begin{equation}
	j_N(t) = \langle \psi_N(t) | \hat{j}_N | \psi_N(t) \rangle \,.
\label{j_as_expectation_value}
\end{equation}
Here,
\begin{equation}
	| \psi_N(t) \rangle = \sum_{m=0}^N c_m e^{-i m^2 t} | m \rangle
\label{psi_N_ket}
\end{equation}
is the particle's state, $| 0 \rangle$, $| 1 \rangle$, $\ldots$, $| N \rangle$  are angular momentum eigenstates satisfying the orthonormality condition $\langle m | n \rangle = \delta_{mn}$, and
\begin{equation}
	\hat{j}_N = \frac{1}{2 \pi} \sum_{m,n=0}^N | m \rangle (m + n) \langle n |
\label{j_N_op}
\end{equation}
is the operator representing the probability current on the subspace of the Hilbert space spanned by $\{ | m \rangle \}_{m=0}^N$. The state $\psi_N$ is assumed to be normalized, i.e. $\langle \psi_N | \psi_N \rangle = 1$, which is equivalent to Eq.~\eqref{norm}.

We now look for eignevectors $| \chi \rangle$ of $\hat{j}_N$ that are of the form
\begin{equation}
	| \chi \rangle = A \sum_{m=0}^N (m + a) | m \rangle \,,
\label{chi_guess}
\end{equation}
where $a$ and $A$ are some (yet to be determined) constants. Substituting Eqs.~\eqref{j_N_op} and \eqref{chi_guess} into the eigenequation
\begin{equation*}
	\hat{j}_N | \chi \rangle = \lambda | \chi \rangle \,,
\end{equation*}
we obtain
\begin{equation*}
	\frac{1}{2 \pi} \sum_{m,n=0}^N (m + n) (n + a) | m \rangle = \lambda \sum_{m=0}^N (m + a) | m \rangle \,.
\end{equation*}
Then, using the identities $\sum_{n=0}^N n = \frac{1}{2} N (N + 1)$ and $\sum_{n=0}^N n^2 = \frac{1}{6} N (N + 1) (2 N + 1)$, we evaluate the sum over $n$ in the last equation to get
\begin{multline*}
	\frac{(N + 1) (N + 2 a)}{4 \pi} \sum_{m=0}^N \left( m + \frac{N (2 N + 1 + 3 a)}{3 (N + 2 a)} \right) | m \rangle \\ = \lambda \sum_{m=0}^N (m + a) | m \rangle \,.
\end{multline*}
The last equation is satisfied if and only if
\begin{equation}
	a = \frac{N (2 N + 1 + 3 a)}{3 (N + 2 a)}
\label{eq_for_a}
\end{equation}
and
\begin{equation}
	\lambda = \frac{(N + 1) (N + 2 a)}{4 \pi} \,.
\label{eq_for_lambda}
\end{equation}
Equation~\eqref{eq_for_a} is quadratic in $a$ and has two roots, $a_+$ and $a_-$, given by
\begin{equation}
	a_{\pm} = \pm \sqrt{\frac{N (2 N + 1)}{6}} \,.
\label{a_pm}	
\end{equation}
Then, according to Eq.~\eqref{eq_for_lambda}, the corresponding eigenvalues, $\lambda_+$ and $\lambda_-$, read
\begin{equation}
	\lambda_{\pm} = \frac{N (N + 1)}{4 \pi} \left( 1 \pm \sqrt{\frac{4 N + 2}{3 N}} \right) \,.
\label{lambda_pm}
\end{equation}
\begin{figure}[h]
	\centering
	\includegraphics[width=0.47\textwidth]{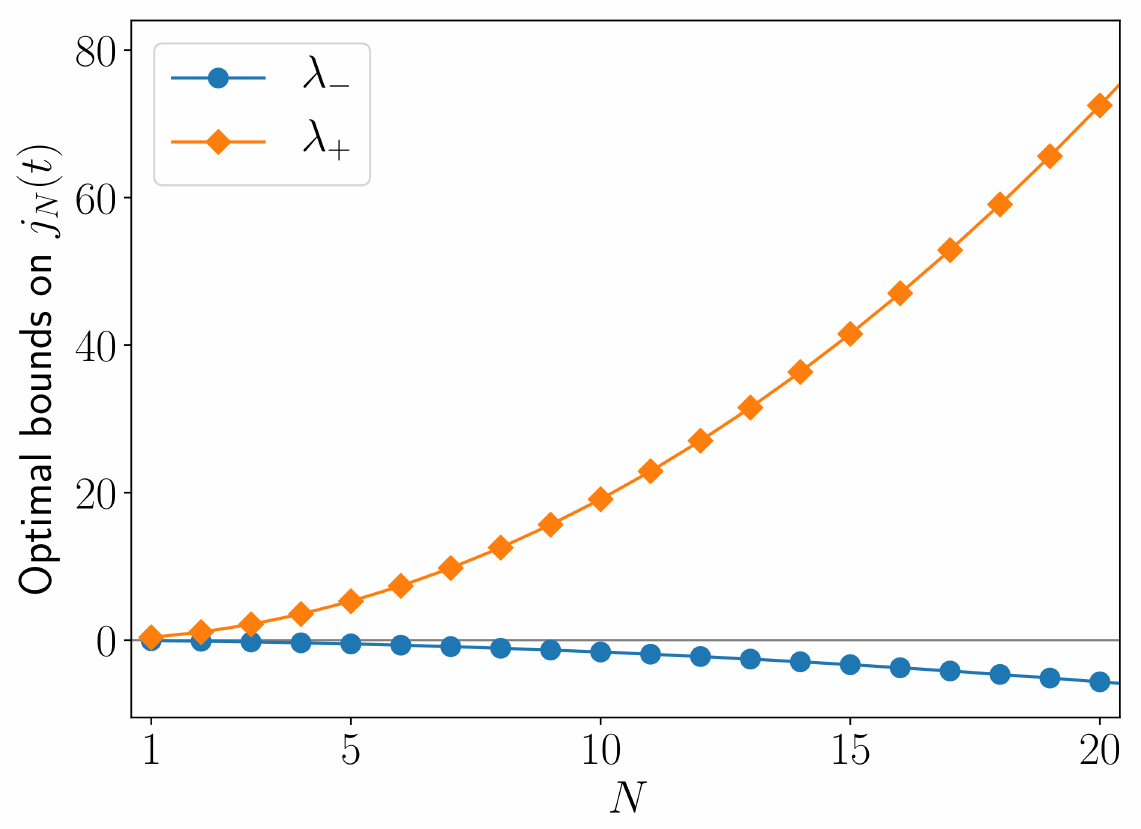}
	\caption{Eigenvalues $\lambda_+$ and $\lambda_-$ as functions of $N$, given by Eq.~\eqref{lambda_pm}. The grey horizontal line shows the level of zero probability current.}
	\label{fig:bounds_on_j}
\end{figure}
Clearly, $\lambda_+ > 0$ and $\lambda_- < 0$, for all $N \ge 1$. Figure~\ref{fig:bounds_on_j} illustrates the dependence of the eigenvalues on $N$. Finally, eigenvectors $| \chi_+ \rangle$ and $| \chi_- \rangle$ corresponding to the eigenvalues $\lambda_+$ and $\lambda_-$ are found by substituting Eq.~\eqref{a_pm} into Eq.~\eqref{chi_guess}:
\begin{equation}
	| \chi_{\pm} \rangle = A_{\pm} \sum_{m=0}^N \left( m \pm \sqrt{\frac{N (2 N + 1)}{6}} \right) | m \rangle \,.
\label{chi_pm}
\end{equation}
It is straightforward to verify (see Appendix~\ref{app:A_pm}) that the normalization constants, $A_+$ and $A_-$ are given by
\begin{equation}
	A_{\pm} = \left[ N (N + 1) \left( \frac{2 N + 1}{3} \pm \sqrt{\frac{N (2 N + 1)}{6}} \right) \right]^{-1/2} .
\label{A_pm}
\end{equation}

We now make the observation that $\lambda_+$ and $\lambda_-$, given by Eq.~\eqref{lambda_pm}, are the only nonzero eigenvalues of the operator $\hat{j}_N$. Indeed, as shown in Appendix~\ref{app:diagonalization_of_j}, $\hat{j}_N$ admits the following decomposition:
\begin{equation}
	\hat{j}_N = \lambda_+ | \chi_+ \rangle \langle \chi_+ | + \lambda_- | \chi_- \rangle \langle \chi_- | \,.
\label{diagonalization_of_j}
\end{equation}
Then, using Eq.~\eqref{j_as_expectation_value}, we get
\begin{equation}
	j_N(t) = \lambda_+ |\langle \chi_+ | \psi_N(t) \rangle|^2 + \lambda_- |\langle \chi_- | \psi_N(t) \rangle|^2 \,.
\label{j_N_decomposition}
\end{equation}	
Since $\lambda_+ > 0$, $\lambda_- < 0$, and $0 \le |\langle \chi_\pm | \psi_N(t) \rangle| \le 1$, we conclude that
\begin{equation*}
	\lambda_- \le j_N(t) \le \lambda_+ \,,
\end{equation*}
which is equivalent to Eq.~\eqref{key_result_1}. The probability current $j_N(t)$ reaches its extreme values, $\lambda_+$ and $\lambda_-$, when $| \psi_N(t) \rangle$ coincides (up to a global phase factor) with the states $| \chi_+ \rangle$ and $| \chi_- \rangle$, respectively.

\section{Probability transfer}
\label{sec:prob_transfer}

We now consider the amount of probability $P_N$ passing through the point $\theta = 0$ over a time interval $\Delta$. More precisely,
\begin{equation*}
	P_N = \int_{-\Delta / 2 }^{\Delta / 2} dT \, J_N(T) \,,
\end{equation*}
where $J_N$ is the probability current defined by Eq.~\eqref{J_N}. In terms of the dimensionless current, defined in Eq.~\eqref{dimensionless}, the probability transfer $P_N$ reads
\begin{equation}
	P_N = \int_{-\alpha}^{\alpha} dt \, j_N(t) \,,
\label{P_N_as_integral}
\end{equation}
where
\begin{equation*}
	\alpha = \frac{\hbar \Delta}{4 M R^2}
\end{equation*}
is a dimensionless parameter.

As demonstrated in Ref.~\cite{Gou21Quantum}, $P_N$ can be negative and has a finite greatest lower bound. In the following section of this paper, we explore the probability current $j_N(t)$ generated by a quantum state characterized by the value of $P_N$ that is very close to the greatest lower bound. To set the stage for this exploration, we now briefly summarize some findings of Ref.~\cite{Gou21Quantum} that are particularly relevant to the ensuing discussion.

The substitution of Eq.~\eqref{j_N} into Eq.~\eqref{P_N_as_integral}, followed by evaluation of the integral over $t$, yields
\begin{equation}
	P_N = \frac{\alpha}{\pi} \sum_{m = 0}^N \sum_{n = 0}^N c^*_m c_n (m + n) \sinc \big[ \alpha (m^2 - n^2) \big] \,,
\label{P_N_as_double_sum}
\end{equation}
where $\sinc z = (\sin z) / z$. To determine the (negative) optimal lower bound for probability transfer, one needs to minimize $P_N$ within the $(N+1)$-dimensional space of vectors $(c_0, c_1, \ldots, c_N)$, while adhering to the normalization constraint \eqref{norm}. This is accomplished through the method of Lagrange multipliers, whereby one conducts unconstrained minimization of the function
\begin{equation*}
	{\cal L}(c_0, c_1, \ldots, c_N) = P_N - \mu \sum_{m=0}^N c_m^* c_m \,,
\end{equation*}
where $\mu$ is a Lagrange multiplier. In view of Eq.~\eqref{P_N_as_double_sum}, the Euler-Lagrange equation corresponding to this minimization problem reads
\begin{equation*}
	\frac{\alpha}{\pi} \sum_{n=0}^N (m + n) \sinc \big[ \alpha (m^2 - n^2) \big] c_n = \mu c_m \,.
\end{equation*}
This equation defines an eigenproblem, in which $\mu$ plays the role of the eigenvalue corresponding to the eigenvector $(c_0, c_1, \ldots, c_N)$.  The eigenproblem is then solved numerically, resulting in a spectrum of $(N+1)$ (not necessarily distinct) eigenvalues, $\mu_0, \mu_1, \ldots, \mu_N$. The smallest eigenvalue corresponds to the desired minimal probability transfer:
\begin{equation*}
	P_N^{\min} \equiv \min P_N = \min \{ \mu_0, \mu_1, \ldots, \mu_N \} \,.
\end{equation*}
It is important to keep in mind that $P_N^{\min}$ depends on the system parameter $\alpha$.

The function $P_N^{\min}$ was numerically computed in Ref.~\cite{Gou21Quantum}. In particular, the study demonstrated that
\begin{equation}
	\inf_{\alpha} \lim_{N \to \infty} P_N^{\min} \simeq -0.116816 \,,
\label{ring_bound}
\end{equation}
thus providing the greatest lower bound on the probability transfer associated with the most general superposition of particle-in-a-ring states with non-negative angular momentum. The bound presented in Eq.~\eqref{ring_bound} is achieved for the system parameter value close to
\begin{equation}
	\alpha = 1.163635 \,.
\label{alpha_min}
\end{equation}

We conclude this section by noting that the probability transfer, $P_N$, can be represented as the sum of two terms, one non-negative and the other non-positive:
\begin{equation}
	P_N = P_N^{(+)} + P_N^{(-)} \,,
\label{P_N_new_representation}
\end{equation}
where
\begin{equation}
	P_N^{(\pm)} = \frac{1}{4 \pi a_{\pm}} \int_{-\alpha}^{\alpha} dt \, \left| \sum_{m=0}^N c_m (m + a_{\pm}) e^{-i m^2 t} \right|^2 \,.
\label{P_N^pm}
\end{equation}
The fact that $a_+ = - a_- > 0$ [see Eq.~\eqref{a_pm}] implies that
\begin{equation*}
	P_N^{(+)} \ge 0 \quad \text{and} \quad P_N^{(-)} \le 0 \,.
\end{equation*} 
Appendix \ref{app:new_representation_of_P_N} provides a derivation of this representation and elucidates its relationship with Eq.~\eqref{P_N_as_double_sum}. Currently, the practical significance of this representation remains unclear. However, the noteworthy aspect that $P_N^{(+)} \ge 0$ and $P_N^{(-)} \le 0$ is nontrivial, and it may prove valuable in future studies, especially when attempting to establish precise bounds for probability transfer.

\section{Fractal dimension of the backflow-maximizing current}
\label{sec:bm_state}

We now explore a specific particle-in-a-ring state with non-negative angular momentum, characterized by a value of probability transfer very close to the estimated bound given in Eq.~\eqref{ring_bound}. The state was obtained via numerical minimization of the probability transfer \eqref{P_N_as_double_sum}, with $N = 9999$ and $\alpha$ given by Eq.~\eqref{alpha_min}, subject to the normalization constraint, Eq.~\eqref{norm}. Employing somewhat loose terminology, we will refer to this state as a ``backflow-maximizing'' state.

More precisely, the backflow-maximizing state $| \psi_{\text{bm}}(t) \rangle$ considered here is defined as $| \psi_{9999}(t) \rangle$ [see Eq.~\eqref{psi_N_ket}] with
\begin{align}
\begin{split}
	&c_0 = 9.443114018508278473 \times 10^{-1} \\
	&c_1 = -3.152130460659169908 \times 10^{-1} \\
	&c_2 = 7.894329104096091398 \times 10^{-2} \\
	& \cdots\cdots\cdots\cdots\cdots\cdots\cdots\cdots\cdots\cdots\cdots\cdots \\
	&c_{9999} = 6.151844737832881660 \times 10^{-10} \,.
\end{split}
\label{c_n_for_bm_state}
\end{align}
The complete list of expansion coefficients is provided in the Supplemental Material \cite{SupplementalMaterial}. The numerically computed value of the probability transfer, corresponding to $| \psi_{\text{bm}} \rangle$, equals $-0.11681564958330892$. Hereinafter, our analysis maintains a fixed value for $\alpha$ as specified by Eq.~\eqref{alpha_min}.

As noted in Ref.~\cite{Gou21Quantum}, the backflow-maximizing state $| \psi_{\text{bm}} \rangle$ is characterized by a finite mean energy, which stands in stark contrast to the infinite mean energy observed in the linear case. In the units used in the present paper, the mean energy of $| \psi_{\text{bm}} \rangle$ is given by
\begin{equation*}
	\langle E \rangle = \frac{\hbar^2}{2 M R^2} \sum_{m=0}^{9999} m^2 c_m^2 \simeq 0.082837 \times \frac{\hbar^2}{M R^2} \,.
\end{equation*}

\begin{figure}[h]
	\centering
	\includegraphics[width=0.49\textwidth]{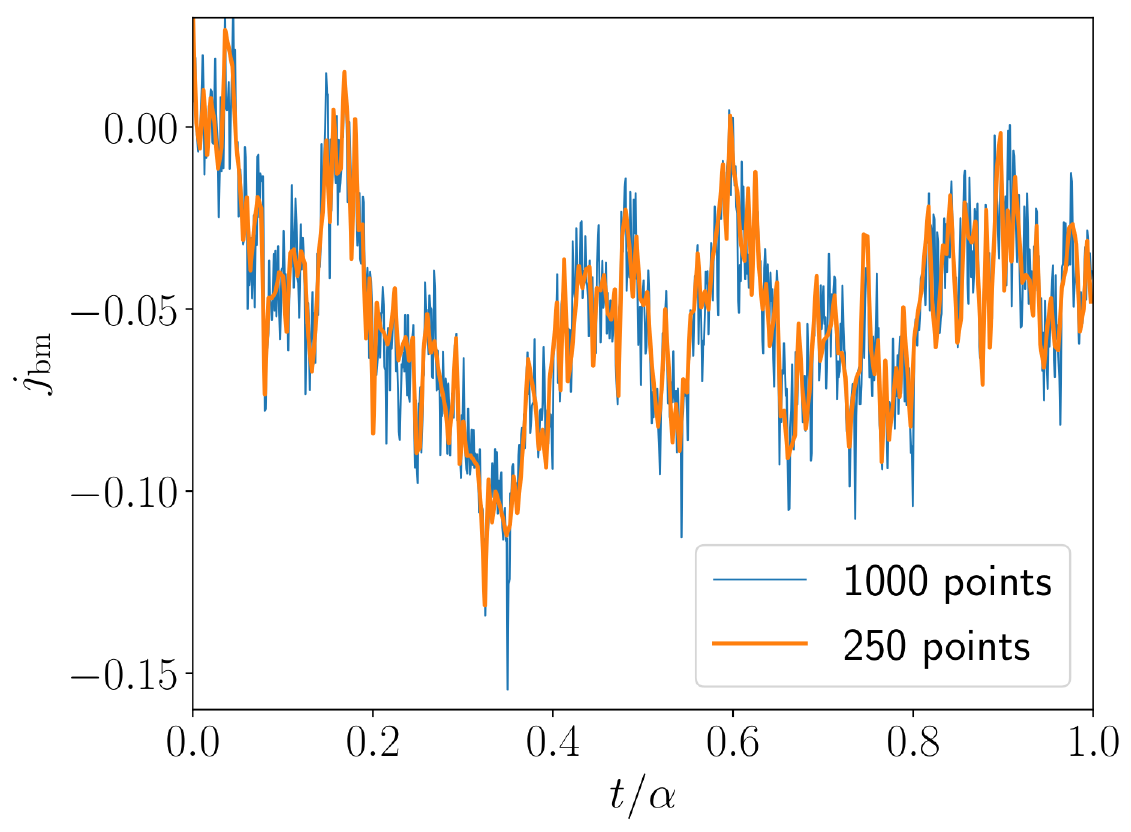}
	\caption{Probability current for the backflow-maximizing state. The two graphs was obtained by computing $j_{9999}(t)$ from Eq.~\eqref{j_N}, with $c_m$'s given by Eq.~\eqref{c_n_for_bm_state}, at two different sets of instances between $t=0$ and $t=\alpha$. 1000 equally spaced points were used to produce the thin (blue) curve, whereas 250 equally spaced points were used for the thick (orange) curve. The value of $\alpha$ is given by Eq.~\eqref{alpha_min}.}
	\label{fig:j_bm_vs_t}
\end{figure}
Figure~\ref{fig:j_bm_vs_t} shows the graph of the probability current $j_{\text{bm}}(t)$, computed numerically for two different sets of points between $t=0$ and $t=\alpha$, using Eq.~\eqref{j_N}, for the backflow-maximizing state, $| \psi_{\text{bm}}(t) \rangle$. In other words, $j_{\text{bm}}(t)$ is $j_{9999}(t)$ [see Eq.~\eqref{j_N}] with $c_m$'s given by Eq.~\eqref{c_n_for_bm_state}. The graph appears to be highly irregular. (In fact, as we will argue below, it represents a fractal curve.) The visual appearance of the graph strongly depends on the particular choice of instances used to evaluate the function $j_{\text{bm}}(t)$ and plot the graph. The thin (blue) curve in Fig.~\ref{fig:j_bm_vs_t} was obtain by computing $j_{\text{bm}}$ at 1000 equally spaced instances between $t=0$ and $t=\alpha$, whereas 250 equally spaced points were used to plot the thick (orange) curve.

The probability current for the backflow-maximizing state of a particle in a ring (see Fig.~\ref{fig:j_bm_vs_t}) is significantly distinct from the corresponding probability current in the particle-on-a-line scenario (see, e.g., Fig.~6 in Ref.~\cite{PGKW06new}). In the latter case, the current is consistently negative within the observed time window for probability transfer, and is represented by a smooth curve. In contrast, the former displays sign changes and exhibits a markedly irregular, fractal-like graph. Exploring this difference is both interesting and important, especially given the proposed direct measurement of current as a strategy for the first experimental observation of quantum backflow \cite{BM94Probability}. To this end, we quantitatively examine the fuzziness and irregularity of the current-versus-time curve for the ring geometry by evaluating its fractal dimension. The remainder of this section focuses on the numerical computation of the fractal dimension, while Sec.~\ref{sec:analytical_approximation} provides analytical arguments supporting the numerical results.

Let us now evaluate the Higuchi dimension \cite{Hig88Approach}, denoted by $D_{\text{H}}$, of the function $j_{\text{bm}}(t)$ on the interval $0 < t < \alpha$ (see Fig.~\ref{fig:j_bm_vs_t}). The Higuchi dimension is widely used as an estimator of the box-counting dimension for bounded functions. (See Ref~\cite{LM20On} for the analysis of the robustness and limitations of this method).

Our calculation of the Higuchi dimension of $j_{\text{bm}}(t)$ follows the procedure presented in Ref.~\cite{Hig88Approach}. We compute the function $j_{\text{bm}}(t)$ at $S = 262144 = 2^{18}$ equally spaced time points between 0 and $\alpha$, thus obtaining the sequence
\begin{equation*}
	{\cal J} = \left\{ {\cal J}(1), {\cal J}(2), {\cal J}(3), \ldots, {\cal J}(S) \right\} \,,
\end{equation*}
where
\begin{equation*}
	{\cal J}(s) = j_{\text{bm}}\left( \frac{s-1}{S-1} \alpha \right) .
\end{equation*}
Then, for an integer $k$ between 1 and $S$, we construct $k$ new sequences:
\begin{widetext}
\begin{align*}
	&{\cal J}_k^{(1)} = \left\{ {\cal J}(1), {\cal J}(1 + k), {\cal J}(1 + 2 k), \ldots, {\cal J} \left( 1 + \left\lfloor \frac{S-1}{k} \right\rfloor k \right) \right\} \\
	&{\cal J}_k^{(2)} = \left\{ {\cal J}(2), {\cal J}(2 + k), {\cal J}(2 + 2 k), \ldots, {\cal J} \left( 2 + \left\lfloor \frac{S-2}{k} \right\rfloor k \right) \right\} \\
	&\cdots\cdots\cdots\cdots\cdots\cdots\cdots\cdots\cdots\cdots\cdots\cdots\cdots\cdots\cdots\cdots\cdots\cdots\cdots\cdots \\
	&{\cal J}_k^{(k)} = \left\{ {\cal J}(k), {\cal J}(k + k), {\cal J}(k + 2 k), \ldots, {\cal J} \left( k + \left\lfloor \frac{S-k}{k} \right\rfloor k \right) \right\}
\end{align*}
\end{widetext}
Here, $\lfloor \cdot \rfloor$ denotes the floor function. For each sequence ${\cal J}_k^{(s)}$, we calculate its ``length''
\begin{equation*}
	L_k^{(s)} = \frac{S-1}{\left\lfloor \frac{S-s}{k} \right\rfloor k^2} \sum_{r=1}^{\left\lfloor \frac{S-s}{k} \right\rfloor} | {\cal J}(s + rk) - {\cal J}(s + (r-1)k) | \,,
\end{equation*}
and then find the average of the lengths:
\begin{equation*}
	L_k = \frac{1}{k} \sum_{s=1}^k L_k^{(s)} \,.
\end{equation*}
If the curve under investigation is a fractal, then $L_k$ scales with $k$ as 
\begin{equation*}
	L_k \sim 1/k^{D_{\text{H}}} \,,
\end{equation*}
with $D_{\text{H}}$ being the Higuchi dimension of the fractal.

\begin{figure}[h]
	\centering
	\includegraphics[width=0.49\textwidth]{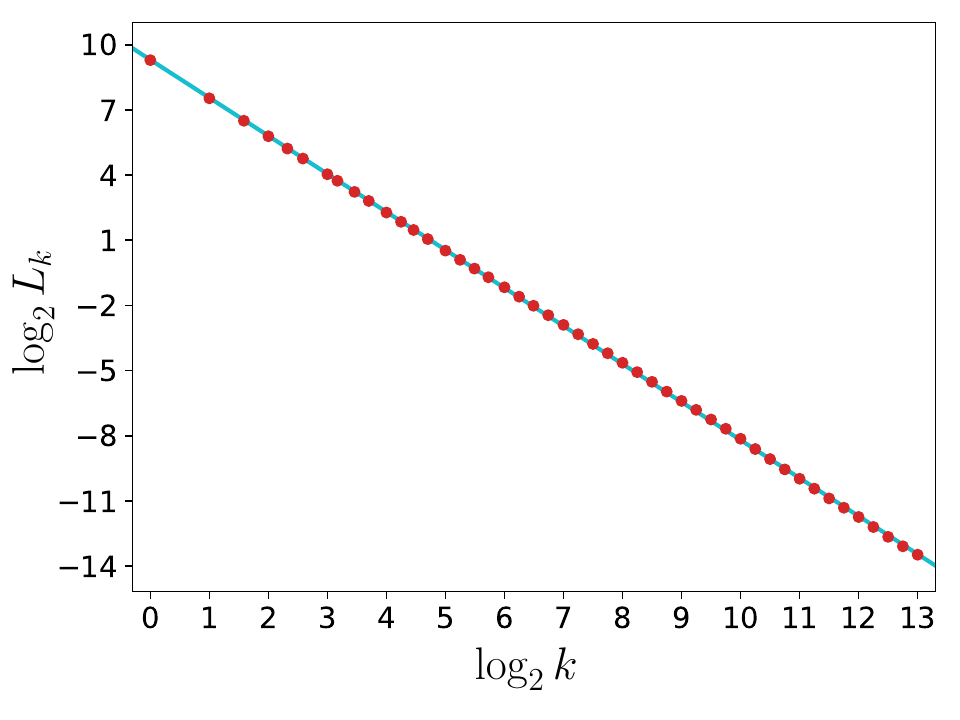}
	\caption{Dependence of $L_k$ on $k$. The (red) circles represent the numerically obtained data points. The (turquoise) line is the corresponding line of best fit. See the text for further discussion.}
	\label{fig:Higuchi}
\end{figure}

Following the above procedure, we calculated $L_k$ for 47 different values of $k$, ranging between 1 and $8192 = 2^{13}$. Figure~\ref{fig:Higuchi} shows the results of this calculations on a log-log scale. The (red) circles represent the 47 data points. The (turquoise) line shows the corresponding line of best fit,
\begin{equation*}
	\log_2 L_k = \text{constant} - D_H \log_2 k \,.
\end{equation*}
The method of least squares gave the following value of the (negative of the) slope of the line:
\begin{equation}
	D_{\text{H}} = 1.751 \pm 0.0019 \,.
\label{D_numerical}
\end{equation}
This is the sought fractal dimension of the probability current of the backflow-maximizing state. (The intercept is of no importance to our analysis. Merely for reference purposes, it approximately equals $9.309$.)

\section{Analytical approximation of the backflow-maximizing state}
\label{sec:analytical_approximation}

In this section, we introduce an accurate analytical approximation for the backflow-maximizing state discussed in the previous section. Termed the ``guess state'', this approximation will be instrumental in exploring the fractal properties of the backflow probability current within the near-optimal regime.

Let us consider the following (guess) state of the particle in a ring:
\begin{equation}
	| \psi_N^{(\text{g})}(t) \rangle = C_N \left( | 0 \rangle - \frac{1}{2} \sum_{m = 1}^N \sinc(\alpha m^2) e^{-i m^2 t} | m \rangle \right) \,,
\label{guess_state}
\end{equation}
where
\begin{equation*}
	C_N = \left( 1 + \frac{1}{4} \sum_{m=1}^N \sinc^2(\alpha m^2) \right)^{-1/2} \,.
\end{equation*}
The state is normalized to unity, $\langle \psi_N^{(\text{g})} | \psi_N^{(\text{g})} \rangle = 1$. As in the previous section, we take the value of $\alpha$ to be specified by Eq.~\eqref{alpha_min}.

For $N=9999$, which is the value used in the numerical investigations in Sec.~\ref{sec:bm_state}, the guess state provides a very good approximation of the backflow-maximizing state, $| \psi_{\text{bm}} \rangle$. The accuracy of the approximation, as quantified by fidelity, exceeds $99\%$:
\begin{equation}
	| \langle \psi_{9999}^{(\text{g})} | \psi_{\text{bm}} \rangle |^2 \simeq 0.996328 \,.
\label{guess_fidelity}
\end{equation}

Let us denote by $j_N^{(\text{g})}(t)$ and $P_N^{(\text{g})}$ the probability current and probability transfer, respectively, associated with the guess state $| \psi_N^{(\text{g})}(t) \rangle$. More precisely,
\begin{equation*}
	j_N^{(\text{g})}(t) = \langle \psi_N^{(\text{g})}(t) | \hat{j}_N | \psi_N^{(\text{g})}(t) \rangle
\end{equation*}
and
\begin{equation*}
	P_N^{(\text{g})} = \int_{-\alpha}^{\alpha} dt \, j_N^{(\text{g})}(t) \,.
\end{equation*}
In terms of numerical calculations, $j_N^{(\text{g})}(t)$ and $P_N^{(\text{g})}$ can be computed from Eqs.~\eqref{j_N} and \eqref{P_N_as_double_sum}, respectively,  by taking the expansion coefficient $c_m$ to be those of $| \psi_N^{(\text{g})}(0) \rangle$, i.e. $c_0 = C_N$, and $c_m = -\frac{C_N}{2} \sinc(\alpha m^2)$ for $1 \le m \le N$. Thus, we find that the guess state $| \psi_{9999}^{(\text{g})} \rangle$ yields the probability transfer
\begin{equation}
	P_{9999}^{(\text{g})} \simeq -0.11131265 \,,
\label{guess_prob_transfer}
\end{equation}
which is over $95\%$ of the bound given by Eq.~\eqref{ring_bound}. For comparison, in the scenario of a particle on a line, the current analytical approximation record for the backflow probability transfer stands at $70\%$ of the BM bound \cite{YHHW12Analytical}.

The numerical estimates provided by Eqs.~\eqref{guess_fidelity} and \eqref{guess_prob_transfer} enable us to infer that the family of guess states $| \psi_N^{(\text{g})} \rangle$, as defined in Eq.~\eqref{guess_state}, serves as a good approximation to the backflow-maximizing state for sufficiently large $N$. Taking this into consideration, we now turn our attention to the limiting state $| \psi_{\infty}^{(\text{g})} \rangle$, given by Eq.~\eqref{guess_state} with $N \to \infty$. We argue that the graph representing the corresponding probability current, $j_{\infty}^{(\text{g})}(t)$, forms a fractal curve with fractal dimension of $\frac{7}{4}$. Our argument relies on the theory outlined in Ref.~\cite{Ber96Quantum}, which can be summarized as follows. Consider a function $f(t)$ defined through the Fourier series
\begin{equation}
	f(t) = \sum_l a_l e^{-i l t} \,.
\label{Fourier}
\end{equation}
If the coefficients $a_l$ have (pseudo)random phases and the power spectrum scales as
\begin{equation*}
	|a_l|^2 \sim \frac{1}{l^{\beta}} \qquad (1 < \beta \le 3)
\end{equation*}
as $|l| \to \infty$, then the graphs of $\real f(t)$ and $\imag f(t)$ are continuous non-differentiable curves with fractal dimension
\begin{equation*}
	D[f] = \frac{5 - \beta}{2} \,.
\end{equation*}

It follows from Eq.~\eqref{j_N} that the probability current in question can be expressed as
\begin{equation*}
	j_{\infty}^{(\text{g})}(t) = \frac{1}{\pi} \real \Big\{ h_0^*(t) h_1(t) \Big\} \,,
\end{equation*}
where
\begin{equation*}
	h_0(t) = \sum_{m=0}^{\infty} c_m e^{-i m^2 t} \,,
\end{equation*}
\begin{equation*}
	h_1(t) = \sum_{m=0}^{\infty} m c_m e^{-i m^2 t} \,,
\end{equation*}
and
\begin{align*}
	&c_0 = C_{\infty} \,, \\
	&c_m = -\frac{C_{\infty}}{2 \alpha} \frac{\sin(\alpha m^2)}{m^2} \qquad (m \ge 1) \,.
\end{align*} 
Both series defining $h_0$ and $h_t$ can be regarded as Fourier series of the form~\eqref{Fourier} with $l = m^2$, and the $\sin(\alpha m^2)$ term furnishes the pseudorandomness of the Fourier coefficients. Then, the power spectrum corresponding to $h_0$ scales as $|a_l|^2 \sim |c_m|^2 \frac{dm}{dl} \sim m^{-5} = l^{-5/2}$, yielding $\beta = 5/2$. Hence, the graph of $h_0(t)$ has fractal dimension
\begin{equation*}
	D[h_0] = \frac{5}{4} \,.
\end{equation*}
In the case of $h_1$, we have $|a_l|^2 \sim |m c_m|^2 \frac{dm}{dl} \sim m^{-3} = l^{-3/2}$, and $\beta = 3/2$. Hence,
\begin{equation*}
	D[h_1] = \frac{7}{4} \,.
\end{equation*}
Given that $j_{\infty}^{(\text{g})}$ is a composite function resulting from the summation of products of two fractal functions, each possessing fractal dimensions of $5/4$ and $7/4$ respectively, it follows that $j_{\infty}^{(\text{g})}$ is itself a fractal, and its fractal dimension is the larger of the two:
\begin{equation}
	D \left[ j_{\infty}^{(\text{g})} \right] = \frac{7}{4} \,.
\label{D_analytical}
\end{equation}

As observed, the guess state, $| \psi_{\infty}^{(g)} \rangle$, serves as a good approximation to the numerically exact backflow-maximizing state, $| \psi_{\text{bm}} \rangle$. Furthermore, the fractal dimension of the probability current linked to the guess state, Eq.~$\eqref{D_analytical}$, aligns closely with the numerical estimate of the fractal dimension characterizing the backflow-maximizing current, Eq.~\eqref{D_numerical}. Hence, it is plausible to conjecture that the probability current embodying the true backflow-maximizing state, if it exists, is indeed a fractal with a dimension of 7/4.

\section{Summary and discussion}
\label{sec:conclusion}

Motivated to gain deeper insights into the phenomenon of quantum backflow for a particle in a ring, we have taken a careful examination of some properties of the time-dependent probability current through a fixed point on the ring. We have shown that when a particle is in a superposition of the $N+1$ lowest energy eigenstates with non-negative angular momentum, Eq.~\eqref{psi_N}, the dimensionless probability current can only range between two extreme values, $\lambda_-$ and $\lambda_+$, Eq.~\eqref{lambda_pm}. $\lambda_-$, being the negative extreme value, determines the limit for an instantaneous measurement of the backflow current. The quantum state $| \chi_- \rangle$ corresponding to this extreme value is given by Eqs.~\eqref{chi_pm} and \eqref{A_pm}.

It is instructive to briefly discuss the regime where $N \gg 1$, wherein the particle state comprises a large number of energy eigenstates. In this regime, the inequalities bounding the dimensionless probability current, given by Eq.~\eqref{key_result_1}, simplify to
\begin{equation*}
	-\frac{\sqrt{\frac{4}{3}}-1}{4 \pi} N^2 \le j_N \le \frac{\sqrt{\frac{4}{3}}+1}{4 \pi} N^2 \,.
\end{equation*}
For the dimensional probability current, Eq.~\eqref{dimensionless}, we have
\begin{equation*}
	-\frac{\sqrt{\frac{4}{3}}-1}{4 \pi} \frac{E_{\max}}{\hbar} \le J_N \le \frac{\sqrt{\frac{4}{3}}+1}{4 \pi} \frac{E_{\max}}{\hbar} \,,
\end{equation*}
where $E_{\max} = \frac{\hbar^2 N^2}{2 M R^2}$ represents the energy of the particle's highest energy component. Presented in this form, our result can be compared with a related statement applicable to the scenario of a free particle on a line \cite{ML00Arrival}: $|J| \le \Delta E / \hbar$, where $J$ represents the probability current and $\Delta E$ denotes the energy uncertainty of the particle's state. While the two statements are not directly analogous, this comparison offers a complementary perspective.

The second part of this study examines the time-dependence of the probability current for particle states that maximize the backflow probability transfer or approach its theoretical bound. First, we perform numerical calculations to determine the backflow-maximizing state (assuming its existence) and compute the fractal dimension of the corresponding probability current versus time function. The obtained numerical value for the fractal dimension, Eq.~\eqref{D_numerical}, falls close to $7/4$. Then, we explore an accurate analytical approximation of the backflow-maximizing state, Eq.~\eqref{guess_state}. This analytical state closely aligns with the numerically computed one, exhibiting a fidelity of over $99\%$ and demonstrating backflow probability transfer exceeding $95\%$ of the theoretical bound. (For comparison, in the scenario of a free particle on a line, the state-of-the-art analytical approximation of the backflow-maximizing state captures approximately $70\%$ of the corresponding maximal probability transfer value \cite{YHHW12Analytical}.) The availability of the accurate analytical approximation enables us to analytically evaluate the fractal dimension of the (almost) backflow-maximizing probability current. The analytically predicted value is $7/4$, consistent with the numerical analysis.

The numerical and analytical findings presented in this study strongly suggest that particle-in-a-ring states approaching the probability transfer bound (of approximately $0.116816$) exhibit fractal characteristics in the time-dependence of the probability current. This observation is not only interesting in its own right but also offers a new signature of quantum backflow. Such a signature could be valuable for future experimental investigations that measure instantaneous current directly (as proposed in Ref.~\cite{BM94Probability}) rather than focusing on integrated probability transfer. Our findings suggest that the instantaneous current will show a highly noisy and irregular time dependence as the system's state approaches backflow-maximizing conditions. This distinctive feature could serve as a guide for exploring the parameter space of the system in search of quantum states that give rise to significant probability backflow.

\appendix
\section{Derivation of Eq.~\eqref{A_pm}}
\label{app:A_pm}

Real constants $A_{\pm}$, normalizing
\begin{equation}
	| \chi_{\pm} \rangle = A_{\pm} \sum_{m=0}^N (m + a_{\pm}) | m \rangle
\label{chi_in_terms_of_A_and_a}
\end{equation}
to unity, are found from the requirement that
\begin{align*}
	1 &= \langle \chi_{\pm} | \chi_{\pm} \rangle \\
	&= A_{\pm}^2 \sum_{m=0}^N (m + a_{\pm})^2 \\
	&= A_{\pm}^2 \left( \sum_{m=0}^N m^2 + 2 a_{\pm} \sum_{m=0}^N m + (N + 1) a_{\pm}^2 \right) \,.
\end{align*}
Using the identities $\sum_{m=0}^N m = \frac{1}{2} N (N + 1)$ and $\sum_{m=0}^N m^2 = \frac{1}{6} N (N + 1) (2 N + 1) = (N + 1) a_{\pm}^2$, we find
\begin{equation}
	A_{\pm} = \left[ (N + 1) (2 a_{\pm}^2 + N a_{\pm}) \right]^{-1/2} \,.
\label{A_in_terms_of_a}
\end{equation}
In view of Eq.~\eqref{a_pm}, the above expression for the normalization constants coincides with the one given by Eq.~\eqref{A_pm}.

\section{Derivation of Eq.~\eqref{diagonalization_of_j}}
\label{app:diagonalization_of_j}

Starting from Eq.~\eqref{chi_in_terms_of_A_and_a}, we rewrite the right-hand side of Eq.~\eqref{diagonalization_of_j} as
\begin{equation*}
	\lambda_+ | \chi_+ \rangle \langle \chi_+ | + \lambda_- | \chi_- \rangle \langle \chi_- | = \sum_{m,n=0}^{\infty} | m \rangle \mathcal{J}_{mn} \langle n | \,, 
\end{equation*}
where
\begin{equation*}
	{\mathcal{J}}_{mn} = \lambda_+ A_+^2 (m + a_+) (n + a_+) + \lambda_- A_-^2 (m + a_-) (n + a_-) \,.
\end{equation*}
Our objective is to establish that $\mathcal{J}_{mn}$ equals $(m + n) / 2 \pi$ [cf.~Eq.~\eqref{j_N_op}]. We have
\begin{align}
\begin{split}
	\mathcal{J}_{mn} &= (\lambda_+ A_+^2 + \lambda_- A_-^2) m n \\ &\quad + (\lambda_+ A_+^2 a_+ + \lambda_- A_-^2 a_-) (m + n) \\ &\quad + \lambda_+ A_+^2 a_+^2 + \lambda_- A_-^2 a_-^2 \,.
\end{split}
\label{J_mn}
\end{align}
Using Eqs.~\eqref{eq_for_lambda} and \eqref{A_in_terms_of_a}, we get
\begin{align}
	\lambda_{\pm} A_{\pm}^2 &= \frac{(N + 1) (N + 2 a_{\pm})}{4 \pi} \frac{1}{(N+1) (2 a_{\pm}^2 + N a_{\pm})} \nonumber \\ &= \frac{1}{4 \pi a_{\pm}} \,. \label{lambda-A-a_relation}
\end{align}
In view of this identity, Eq.~\eqref{J_mn} becomes
\begin{equation*}
	\mathcal{J}_{mn} = \frac{1}{4 \pi} \left( \frac{1}{a_+} + \frac{1}{a_-} \right) m n + \frac{1}{2 \pi} (m + n) + \frac{a_+ + a_-}{4 \pi} \,.
\end{equation*}
Finally, using fact that $a_+ = -a_-$ [see Eq.~\eqref{a_pm}], we arrive at the sought result:
\begin{equation*}
	\mathcal{J}_{mn} = \frac{m + n}{2 \pi} \,.
\end{equation*}

\section{Derivation of Eqs.~\eqref{P_N_new_representation} and \eqref{P_N^pm}}
\label{app:new_representation_of_P_N}

Substituting the diagonal representation of the probability current, given by Eq.~\eqref{j_N_decomposition}, into Eq.~\eqref{P_N_as_integral}, we get
\begin{equation*}
	P_N = \lambda_+ \int_{-\alpha}^{\alpha} dt \, |\langle \chi_+ | \psi_N(t) \rangle|^2 + \lambda_- \int_{-\alpha}^{\alpha} dt \, |\langle \chi_- | \psi_N(t) \rangle|^2 \,.
\end{equation*}
In view of Eqs.~\eqref{psi_N_ket} and \eqref{chi_in_terms_of_A_and_a}, we have
\begin{equation*}
	\langle \chi_{\pm} | \psi_N(t) \rangle = A_{\pm} \sum_{m=0}^N c_m (m + a_{\pm}) e^{-i m^2 t} \,,
\end{equation*}
so that
\begin{multline*}
	P_N = \lambda_+ A_+^2 \int_{-\alpha}^{\alpha} dt \, \left| \sum_{m=0}^N c_m (m + a_+) e^{-i m^2 t} \right|^2 \\ + \lambda_- A_-^2 \int_{-\alpha}^{\alpha} dt \, \left| \sum_{m=0}^N c_m (m + a_-) e^{-i m^2 t} \right|^2 .
\end{multline*}
Then, taking into account Eq.~\eqref{lambda-A-a_relation}, we arrive at
\begin{align*}
	P_N &= \frac{1}{4 \pi a_+} \int_{-\alpha}^{\alpha} dt \, \left| \sum_{m=0}^N c_m (m + a_+) e^{-i m^2 t} \right|^2 \\ &\qquad + \frac{1}{4 \pi a_-} \int_{-\alpha}^{\alpha} dt \, \left| \sum_{m=0}^N c_m (m + a_-) e^{-i m^2 t} \right|^2 \\
	&= P_N^{(+)} + P_N^{(-)} \,,
\end{align*}
which is the representation given by Eqs.~\eqref{P_N_new_representation} and \eqref{P_N^pm}.

In order to better understand the connection between the last representation and the one given by Eq.~\eqref{P_N_as_double_sum}, let us perform the integration over $t$ explicitly and demonstrate that the non-negative and non-positive components of the probability transfer indeed add up to the value given by Eq.~\eqref{P_N_as_double_sum}. We have
\begin{align*}
	&P_N^{(\pm)} \\ &= \frac{1}{4 \pi a_{\pm}} \sum_{m,n=0}^N c_m^* c_n (m + a_{\pm}) (n + a_{\pm}) \int_{-\alpha}^{\alpha} dt \, e^{i (m^2 - n^2) t} \\
	&= \frac{\alpha}{2 \pi a_{\pm}} \sum_{m,n=0}^N c_m^* c_n (m + a_{\pm}) (n + a_{\pm}) \sinc \big[ \alpha (m^2 - n^2) \big] \,.
\end{align*}
Then,
\begin{equation*}
	P_N^{(+)} + P_N^{(-)} = \frac{\alpha}{\pi} \sum_{m = 0}^N \sum_{n = 0}^N c^*_m c_n S_{mn} \sinc \big[ \alpha (m^2 - n^2) \big] \,,
\end{equation*}
where
\begin{equation*}
	S_{mn} = \frac{(m + a_+) (n + a_+)}{2 a_+} + \frac{(m + a_-) (n + a_-)}{2 a_-} \,.
\end{equation*}
Finally, using the fact that $a_- = - a_+$ [see Eq.~\eqref{a_pm}], we obtain
\begin{align*}
	S_{mn} &= \frac{(m + a_+) (n + a_+)}{2 a_+} - \frac{(m - a_+) (n - a_+)}{2 a_+} \\
	&= m + n \,.
\end{align*}
This implies that
\begin{equation*}
P_N^{(+)} + P_N^{(-)} = \frac{\alpha}{\pi} \sum_{m = 0}^N \sum_{n = 0}^N c^*_m c_n (m + n) \sinc \big[ \alpha (m^2 - n^2) \big] \,.
\end{equation*}
The expression in the right-hand side of this equality coincides with the one in the right-hand side of Eq.~\eqref{P_N_as_double_sum}.

%%%%%%%%%%%%%%%%%%%%%%%%%%%%%%%%%%%%%%%%%%%%%%%%%%%%%%%%%%%%%%%%%%%%%%
%


\begin{thebibliography}{55}%
\makeatletter
\providecommand \@ifxundefined [1]{%
 \@ifx{#1\undefined}
}%
\providecommand \@ifnum [1]{%
 \ifnum #1\expandafter \@firstoftwo
 \else \expandafter \@secondoftwo
 \fi
}%
\providecommand \@ifx [1]{%
 \ifx #1\expandafter \@firstoftwo
 \else \expandafter \@secondoftwo
 \fi
}%
\providecommand \natexlab [1]{#1}%
\providecommand \enquote  [1]{``#1''}%
\providecommand \bibnamefont  [1]{#1}%
\providecommand \bibfnamefont [1]{#1}%
\providecommand \citenamefont [1]{#1}%
\providecommand \href@noop [0]{\@secondoftwo}%
\providecommand \href [0]{\begingroup \@sanitize@url \@href}%
\providecommand \@href[1]{\@@startlink{#1}\@@href}%
\providecommand \@@href[1]{\endgroup#1\@@endlink}%
\providecommand \@sanitize@url [0]{\catcode `\\12\catcode `\$12\catcode
  `\&12\catcode `\#12\catcode `\^12\catcode `\_12\catcode `\%12\relax}%
\providecommand \@@startlink[1]{}%
\providecommand \@@endlink[0]{}%
\providecommand \url  [0]{\begingroup\@sanitize@url \@url }%
\providecommand \@url [1]{\endgroup\@href {#1}{\urlprefix }}%
\providecommand \urlprefix  [0]{URL }%
\providecommand \Eprint [0]{\href }%
\providecommand \doibase [0]{http://dx.doi.org/}%
\providecommand \selectlanguage [0]{\@gobble}%
\providecommand \bibinfo  [0]{\@secondoftwo}%
\providecommand \bibfield  [0]{\@secondoftwo}%
\providecommand \translation [1]{[#1]}%
\providecommand \BibitemOpen [0]{}%
\providecommand \bibitemStop [0]{}%
\providecommand \bibitemNoStop [0]{.\EOS\space}%
\providecommand \EOS [0]{\spacefactor3000\relax}%
\providecommand \BibitemShut  [1]{\csname bibitem#1\endcsname}%
\let\auto@bib@innerbib\@empty
%</preamble>
\bibitem [{\citenamefont {Allcock}(1969)}]{All69time-c}%
  \BibitemOpen
  \bibfield  {author} {\bibinfo {author} {\bibfnamefont {G.~R.}\ \bibnamefont
  {Allcock}},\ }\bibfield  {title} {\enquote {\bibinfo {title} {{The time of
  arrival in quantum mechanics III. The measurement ensemble}},}\ }\href
  {\doibase 10.1016/0003-4916(69)90253-X} {\bibfield  {journal} {\bibinfo
  {journal} {Ann. Phys. (N. Y).}\ }\textbf {\bibinfo {volume} {53}},\ \bibinfo
  {pages} {311} (\bibinfo {year} {1969})}\BibitemShut {NoStop}%
\bibitem [{\citenamefont {Kijowski}(1974)}]{Kij74time}%
  \BibitemOpen
  \bibfield  {author} {\bibinfo {author} {\bibfnamefont {J.}~\bibnamefont
  {Kijowski}},\ }\bibfield  {title} {\enquote {\bibinfo {title} {{On the time
  operator in quantum mechanics and the Heisenberg uncertainty relation for
  energy and time}},}\ }\href {\doibase 10.1016/S0034-4877(74)80004-2}
  {\bibfield  {journal} {\bibinfo  {journal} {Rep. Math. Phys.}\ }\textbf
  {\bibinfo {volume} {6}},\ \bibinfo {pages} {361} (\bibinfo {year}
  {1974})}\BibitemShut {NoStop}%
\bibitem [{\citenamefont {Werner}(1988)}]{Wer88Wigner}%
  \BibitemOpen
  \bibfield  {author} {\bibinfo {author} {\bibfnamefont {R.~F.}\ \bibnamefont
  {Werner}},\ }\bibfield  {title} {\enquote {\bibinfo {title} {{Wigner
  quantisation of arrival time and oscillator phase}},}\ }\href {\doibase
  10.1088/0305-4470/21/24/012} {\bibfield  {journal} {\bibinfo  {journal} {J.
  Phys. A: Math. Gen.}\ }\textbf {\bibinfo {volume} {21}},\ \bibinfo {pages}
  {4565} (\bibinfo {year} {1988})}\BibitemShut {NoStop}%
\bibitem [{\citenamefont {Bracken}\ and\ \citenamefont
  {Melloy}(1994)}]{BM94Probability}%
  \BibitemOpen
  \bibfield  {author} {\bibinfo {author} {\bibfnamefont {A.~J.}\ \bibnamefont
  {Bracken}}\ and\ \bibinfo {author} {\bibfnamefont {G.~F.}\ \bibnamefont
  {Melloy}},\ }\bibfield  {title} {\enquote {\bibinfo {title} {{Probability
  backflow and a new dimensionless quantum number}},}\ }\href {\doibase
  10.1088/0305-4470/27/6/040} {\bibfield  {journal} {\bibinfo  {journal} {J.
  Phys. A: Math. Gen.}\ }\textbf {\bibinfo {volume} {27}},\ \bibinfo {pages}
  {2197} (\bibinfo {year} {1994})}\BibitemShut {NoStop}%
\bibitem [{\citenamefont {Eveson}\ \emph {et~al.}(2005)\citenamefont {Eveson},
  \citenamefont {Fewster},\ and\ \citenamefont {Verch}}]{EFV05Quantum}%
  \BibitemOpen
  \bibfield  {author} {\bibinfo {author} {\bibfnamefont {S.~P.}\ \bibnamefont
  {Eveson}}, \bibinfo {author} {\bibfnamefont {C.~J.}\ \bibnamefont {Fewster}},
  \ and\ \bibinfo {author} {\bibfnamefont {R.}~\bibnamefont {Verch}},\
  }\bibfield  {title} {\enquote {\bibinfo {title} {{Quantum Inequalities in
  Quantum Mechanics}},}\ }\href {\doibase 10.1007/s00023-005-0197-9} {\bibfield
   {journal} {\bibinfo  {journal} {Ann. Henri Poincar\'e}\ }\textbf {\bibinfo
  {volume} {6}},\ \bibinfo {pages} {1} (\bibinfo {year} {2005})}\BibitemShut
  {NoStop}%
\bibitem [{\citenamefont {Penz}\ \emph {et~al.}(2006)\citenamefont {Penz},
  \citenamefont {Gr{\"{u}}bl}, \citenamefont {Kreidl},\ and\ \citenamefont
  {Wagner}}]{PGKW06new}%
  \BibitemOpen
  \bibfield  {author} {\bibinfo {author} {\bibfnamefont {M.}~\bibnamefont
  {Penz}}, \bibinfo {author} {\bibfnamefont {G.}~\bibnamefont {Gr{\"{u}}bl}},
  \bibinfo {author} {\bibfnamefont {S.}~\bibnamefont {Kreidl}}, \ and\ \bibinfo
  {author} {\bibfnamefont {P.}~\bibnamefont {Wagner}},\ }\bibfield  {title}
  {\enquote {\bibinfo {title} {{A new approach to quantum backflow}},}\ }\href
  {\doibase 10.1088/0305-4470/39/2/012} {\bibfield  {journal} {\bibinfo
  {journal} {J. Phys. A: Math. Gen.}\ }\textbf {\bibinfo {volume} {39}},\
  \bibinfo {pages} {423} (\bibinfo {year} {2006})}\BibitemShut {NoStop}%
\bibitem [{\citenamefont {Trillo}\ \emph {et~al.}(2023)\citenamefont {Trillo},
  \citenamefont {Le},\ and\ \citenamefont {Navascu{\'{e}}s}}]{TLN23Quantum}%
  \BibitemOpen
  \bibfield  {author} {\bibinfo {author} {\bibfnamefont {D.}~\bibnamefont
  {Trillo}}, \bibinfo {author} {\bibfnamefont {T.~P.}\ \bibnamefont {Le}}, \
  and\ \bibinfo {author} {\bibfnamefont {M.}~\bibnamefont {Navascu{\'{e}}s}},\
  }\bibfield  {title} {\enquote {\bibinfo {title} {{Quantum advantages for
  transportation tasks - projectiles, rockets and quantum backflow}},}\ }\href
  {\doibase 10.1038/s41534-023-00739-z} {\bibfield  {journal} {\bibinfo
  {journal} {npj Quantum Inf.}\ }\textbf {\bibinfo {volume} {9}},\ \bibinfo
  {pages} {69} (\bibinfo {year} {2023})}\BibitemShut {NoStop}%
\bibitem [{\citenamefont {Melloy}\ and\ \citenamefont
  {Bracken}(1998{\natexlab{a}})}]{MB98velocity}%
  \BibitemOpen
  \bibfield  {author} {\bibinfo {author} {\bibfnamefont {G.~F.}\ \bibnamefont
  {Melloy}}\ and\ \bibinfo {author} {\bibfnamefont {A.~J.}\ \bibnamefont
  {Bracken}},\ }\bibfield  {title} {\enquote {\bibinfo {title} {{The velocity
  of probability transport in quantum mechanics}},}\ }\href {\doibase
  10.1002/andp.199851007-818} {\bibfield  {journal} {\bibinfo  {journal} {Ann.
  Phys. (Berlin, Ger.)}\ }\textbf {\bibinfo {volume} {510}},\ \bibinfo {pages}
  {726} (\bibinfo {year} {1998}{\natexlab{a}})}\BibitemShut {NoStop}%
\bibitem [{\citenamefont {Berry}(2010)}]{Ber10Quantum}%
  \BibitemOpen
  \bibfield  {author} {\bibinfo {author} {\bibfnamefont {M.~V.}\ \bibnamefont
  {Berry}},\ }\bibfield  {title} {\enquote {\bibinfo {title} {{Quantum
  backflow, negative kinetic energy, and optical retro-propagation}},}\ }\href
  {\doibase 10.1088/1751-8113/43/41/415302} {\bibfield  {journal} {\bibinfo
  {journal} {J. Phys. A: Math. Theor.}\ }\textbf {\bibinfo {volume} {43}},\
  \bibinfo {pages} {415302} (\bibinfo {year} {2010})}\BibitemShut {NoStop}%
\bibitem [{\citenamefont {Bostelmann}\ \emph {et~al.}(2017)\citenamefont
  {Bostelmann}, \citenamefont {Cadamuro},\ and\ \citenamefont
  {Lechner}}]{BCL17Quantum}%
  \BibitemOpen
  \bibfield  {author} {\bibinfo {author} {\bibfnamefont {H.}~\bibnamefont
  {Bostelmann}}, \bibinfo {author} {\bibfnamefont {D.}~\bibnamefont
  {Cadamuro}}, \ and\ \bibinfo {author} {\bibfnamefont {G.}~\bibnamefont
  {Lechner}},\ }\bibfield  {title} {\enquote {\bibinfo {title} {{Quantum
  backflow and scattering}},}\ }\href {\doibase 10.1103/PhysRevA.96.012112}
  {\bibfield  {journal} {\bibinfo  {journal} {Phys. Rev. A}\ }\textbf {\bibinfo
  {volume} {96}},\ \bibinfo {pages} {012112} (\bibinfo {year}
  {2017})}\BibitemShut {NoStop}%
\bibitem [{\citenamefont {Muga}\ \emph {et~al.}(1999)\citenamefont {Muga},
  \citenamefont {Palao},\ and\ \citenamefont {Leavens}}]{MPL99Arrival}%
  \BibitemOpen
  \bibfield  {author} {\bibinfo {author} {\bibfnamefont {J.~G.}\ \bibnamefont
  {Muga}}, \bibinfo {author} {\bibfnamefont {J.~P.}\ \bibnamefont {Palao}}, \
  and\ \bibinfo {author} {\bibfnamefont {C.~R.}\ \bibnamefont {Leavens}},\
  }\bibfield  {title} {\enquote {\bibinfo {title} {{Arrival time distributions
  and perfect absorption in classical and quantum mechanics}},}\ }\href
  {\doibase 10.1016/S0375-9601(99)00020-1} {\bibfield  {journal} {\bibinfo
  {journal} {Phys. Lett. A}\ }\textbf {\bibinfo {volume} {253}},\ \bibinfo
  {pages} {21} (\bibinfo {year} {1999})}\BibitemShut {NoStop}%
\bibitem [{\citenamefont {Muga}\ and\ \citenamefont
  {Leavens}(2000)}]{ML00Arrival}%
  \BibitemOpen
  \bibfield  {author} {\bibinfo {author} {\bibfnamefont {J.~G.}\ \bibnamefont
  {Muga}}\ and\ \bibinfo {author} {\bibfnamefont {C.~R.}\ \bibnamefont
  {Leavens}},\ }\bibfield  {title} {\enquote {\bibinfo {title} {{Arrival time
  in quantum mechanics}},}\ }\href {\doibase 10.1016/S0370-1573(00)00047-8}
  {\bibfield  {journal} {\bibinfo  {journal} {Phys. Rep.}\ }\textbf {\bibinfo
  {volume} {338}},\ \bibinfo {pages} {353} (\bibinfo {year}
  {2000})}\BibitemShut {NoStop}%
\bibitem [{\citenamefont {Damborenea}\ \emph {et~al.}(2002)\citenamefont
  {Damborenea}, \citenamefont {Egusquiza}, \citenamefont {Hegerfeldt},\ and\
  \citenamefont {Muga}}]{DEHM02Measurement}%
  \BibitemOpen
  \bibfield  {author} {\bibinfo {author} {\bibfnamefont {J.~A.}\ \bibnamefont
  {Damborenea}}, \bibinfo {author} {\bibfnamefont {I.~L.}\ \bibnamefont
  {Egusquiza}}, \bibinfo {author} {\bibfnamefont {G.~C.}\ \bibnamefont
  {Hegerfeldt}}, \ and\ \bibinfo {author} {\bibfnamefont {J.~G.}\ \bibnamefont
  {Muga}},\ }\bibfield  {title} {\enquote {\bibinfo {title} {{Measurement-based
  approach to quantum arrival times}},}\ }\href {\doibase
  10.1103/PhysRevA.66.052104} {\bibfield  {journal} {\bibinfo  {journal} {Phys.
  Rev. A}\ }\textbf {\bibinfo {volume} {66}},\ \bibinfo {pages} {052104}
  (\bibinfo {year} {2002})}\BibitemShut {NoStop}%
\bibitem [{\citenamefont {Halliwell}\ \emph {et~al.}(2019)\citenamefont
  {Halliwell}, \citenamefont {Beck}, \citenamefont {Lee},\ and\ \citenamefont
  {O'Brien}}]{HBLO19Quasiprobability}%
  \BibitemOpen
  \bibfield  {author} {\bibinfo {author} {\bibfnamefont {J.~J.}\ \bibnamefont
  {Halliwell}}, \bibinfo {author} {\bibfnamefont {H.}~\bibnamefont {Beck}},
  \bibinfo {author} {\bibfnamefont {B.~K.~B.}\ \bibnamefont {Lee}}, \ and\
  \bibinfo {author} {\bibfnamefont {S.}~\bibnamefont {O'Brien}},\ }\bibfield
  {title} {\enquote {\bibinfo {title} {{Quasiprobability for the arrival-time
  problem with links to backflow and the Leggett-Garg inequalities}},}\ }\href
  {\doibase 10.1103/PhysRevA.99.012124} {\bibfield  {journal} {\bibinfo
  {journal} {Phys. Rev. A}\ }\textbf {\bibinfo {volume} {99}},\ \bibinfo
  {pages} {012124} (\bibinfo {year} {2019})}\BibitemShut {NoStop}%
\bibitem [{\citenamefont {Strange}(2012)}]{Str12Large}%
  \BibitemOpen
  \bibfield  {author} {\bibinfo {author} {\bibfnamefont {P.}~\bibnamefont
  {Strange}},\ }\bibfield  {title} {\enquote {\bibinfo {title} {{Large quantum
  probability backflow and the azimuthal angle–angular momentum uncertainty
  relation for an electron in a constant magnetic field}},}\ }\href {\doibase
  10.1088/0143-0807/33/5/1147} {\bibfield  {journal} {\bibinfo  {journal} {Eur.
  J. Phys.}\ }\textbf {\bibinfo {volume} {33}},\ \bibinfo {pages} {1147}
  (\bibinfo {year} {2012})}\BibitemShut {NoStop}%
\bibitem [{\citenamefont {Paccoia}\ \emph {et~al.}(2020)\citenamefont
  {Paccoia}, \citenamefont {Panella},\ and\ \citenamefont
  {Roy}}]{PPR20Angular}%
  \BibitemOpen
  \bibfield  {author} {\bibinfo {author} {\bibfnamefont {V.~D.}\ \bibnamefont
  {Paccoia}}, \bibinfo {author} {\bibfnamefont {O.}~\bibnamefont {Panella}}, \
  and\ \bibinfo {author} {\bibfnamefont {P.}~\bibnamefont {Roy}},\ }\bibfield
  {title} {\enquote {\bibinfo {title} {Angular momentum quantum backflow in the
  noncommutative plane},}\ }\href {\doibase 10.1103/PhysRevA.102.062218}
  {\bibfield  {journal} {\bibinfo  {journal} {Phys. Rev. A}\ }\textbf {\bibinfo
  {volume} {102}},\ \bibinfo {pages} {062218} (\bibinfo {year}
  {2020})}\BibitemShut {NoStop}%
\bibitem [{\citenamefont {Goussev}(2021)}]{Gou21Quantum}%
  \BibitemOpen
  \bibfield  {author} {\bibinfo {author} {\bibfnamefont {A.}~\bibnamefont
  {Goussev}},\ }\bibfield  {title} {\enquote {\bibinfo {title} {{Quantum
  backflow in a ring}},}\ }\href {\doibase 10.1103/PhysRevA.103.022217}
  {\bibfield  {journal} {\bibinfo  {journal} {Phys. Rev. A}\ }\textbf {\bibinfo
  {volume} {103}},\ \bibinfo {pages} {022217} (\bibinfo {year}
  {2021})}\BibitemShut {NoStop}%
\bibitem [{\citenamefont {Barbier}\ \emph
  {et~al.}(2023{\natexlab{a}})\citenamefont {Barbier}, \citenamefont
  {Goussev},\ and\ \citenamefont {Srivastava}}]{BGS23Unbounded}%
  \BibitemOpen
  \bibfield  {author} {\bibinfo {author} {\bibfnamefont {M.}~\bibnamefont
  {Barbier}}, \bibinfo {author} {\bibfnamefont {A.}~\bibnamefont {Goussev}}, \
  and\ \bibinfo {author} {\bibfnamefont {S.~C.~L.}\ \bibnamefont
  {Srivastava}},\ }\bibfield  {title} {\enquote {\bibinfo {title} {{Unbounded
  quantum backflow in two dimensions}},}\ }\href {\doibase
  10.1103/PhysRevA.107.032204} {\bibfield  {journal} {\bibinfo  {journal}
  {Phys. Rev. A}\ }\textbf {\bibinfo {volume} {107}},\ \bibinfo {pages}
  {032204} (\bibinfo {year} {2023}{\natexlab{a}})}\BibitemShut {NoStop}%
\bibitem [{\citenamefont {Barbier}(2020)}]{Bar20Quantum}%
  \BibitemOpen
  \bibfield  {author} {\bibinfo {author} {\bibfnamefont {M.}~\bibnamefont
  {Barbier}},\ }\bibfield  {title} {\enquote {\bibinfo {title} {{Quantum
  backflow for many-particle systems}},}\ }\href {\doibase
  10.1103/PhysRevA.102.023334} {\bibfield  {journal} {\bibinfo  {journal}
  {Phys. Rev. A}\ }\textbf {\bibinfo {volume} {102}},\ \bibinfo {pages}
  {023334} (\bibinfo {year} {2020})}\BibitemShut {NoStop}%
\bibitem [{\citenamefont {Mousavi}\ and\ \citenamefont
  {Miret-Art{\'{e}}s}(2020{\natexlab{a}})}]{MM20Quantum}%
  \BibitemOpen
  \bibfield  {author} {\bibinfo {author} {\bibfnamefont {S.~V.}\ \bibnamefont
  {Mousavi}}\ and\ \bibinfo {author} {\bibfnamefont {S.}~\bibnamefont
  {Miret-Art{\'{e}}s}},\ }\bibfield  {title} {\enquote {\bibinfo {title}
  {{Quantum backflow for dissipative two-identical-particle systems}},}\ }\href
  {\doibase 10.1016/j.rinp.2020.103426} {\bibfield  {journal} {\bibinfo
  {journal} {Results Phys.}\ }\textbf {\bibinfo {volume} {19}},\ \bibinfo
  {pages} {103426} (\bibinfo {year} {2020}{\natexlab{a}})}\BibitemShut
  {NoStop}%
\bibitem [{\citenamefont {Albarelli}\ \emph {et~al.}(2016)\citenamefont
  {Albarelli}, \citenamefont {Guaita},\ and\ \citenamefont
  {Paris}}]{AGP16Quantum}%
  \BibitemOpen
  \bibfield  {author} {\bibinfo {author} {\bibfnamefont {F.}~\bibnamefont
  {Albarelli}}, \bibinfo {author} {\bibfnamefont {T.}~\bibnamefont {Guaita}}, \
  and\ \bibinfo {author} {\bibfnamefont {M.~G.~A.}\ \bibnamefont {Paris}},\
  }\bibfield  {title} {\enquote {\bibinfo {title} {{Quantum backflow effect and
  nonclassicality}},}\ }\href {\doibase 10.1142/S0219749916500325} {\bibfield
  {journal} {\bibinfo  {journal} {Int. J. Quantum Inf.}\ }\textbf {\bibinfo
  {volume} {14}},\ \bibinfo {pages} {1650032} (\bibinfo {year}
  {2016})}\BibitemShut {NoStop}%
\bibitem [{\citenamefont {Mousavi}\ and\ \citenamefont
  {Miret-Art{\'{e}}s}(2020{\natexlab{b}})}]{MM20Dissipative}%
  \BibitemOpen
  \bibfield  {author} {\bibinfo {author} {\bibfnamefont {S.~V.}\ \bibnamefont
  {Mousavi}}\ and\ \bibinfo {author} {\bibfnamefont {S.}~\bibnamefont
  {Miret-Art{\'{e}}s}},\ }\bibfield  {title} {\enquote {\bibinfo {title}
  {{Dissipative quantum backflow}},}\ }\href {\doibase
  10.1140/epjp/s13360-020-00336-5} {\bibfield  {journal} {\bibinfo  {journal}
  {Eur. Phys. J. Plus}\ }\textbf {\bibinfo {volume} {135}},\ \bibinfo {pages}
  {324} (\bibinfo {year} {2020}{\natexlab{b}})}\BibitemShut {NoStop}%
\bibitem [{\citenamefont {Mousavi}\ and\ \citenamefont
  {Miret-Art{\'{e}}s}(2020{\natexlab{c}})}]{MM20Erratum}%
  \BibitemOpen
  \bibfield  {author} {\bibinfo {author} {\bibfnamefont {S.~V.}\ \bibnamefont
  {Mousavi}}\ and\ \bibinfo {author} {\bibfnamefont {S.}~\bibnamefont
  {Miret-Art{\'{e}}s}},\ }\bibfield  {title} {\enquote {\bibinfo {title}
  {{Erratum to: Dissipative quantum backflow}},}\ }\href {\doibase
  10.1140/epjp/s13360-020-00655-7} {\bibfield  {journal} {\bibinfo  {journal}
  {Eur. Phys. J. Plus}\ }\textbf {\bibinfo {volume} {135}},\ \bibinfo {pages}
  {654} (\bibinfo {year} {2020}{\natexlab{c}})}\BibitemShut {NoStop}%
\bibitem [{\citenamefont {Melloy}\ and\ \citenamefont
  {Bracken}(1998{\natexlab{b}})}]{MB98Probability}%
  \BibitemOpen
  \bibfield  {author} {\bibinfo {author} {\bibfnamefont {G.~F.}\ \bibnamefont
  {Melloy}}\ and\ \bibinfo {author} {\bibfnamefont {A.~J.}\ \bibnamefont
  {Bracken}},\ }\bibfield  {title} {\enquote {\bibinfo {title} {{Probability
  Backflow for a Dirac Particle}},}\ }\href {\doibase 10.1023/A:1018724313788}
  {\bibfield  {journal} {\bibinfo  {journal} {Found. Phys.}\ }\textbf {\bibinfo
  {volume} {28}},\ \bibinfo {pages} {505} (\bibinfo {year}
  {1998}{\natexlab{b}})}\BibitemShut {NoStop}%
\bibitem [{\citenamefont {Su}\ and\ \citenamefont {Chen}(2018)}]{SC18Quantum}%
  \BibitemOpen
  \bibfield  {author} {\bibinfo {author} {\bibfnamefont {H.}~\bibnamefont
  {Su}}\ and\ \bibinfo {author} {\bibfnamefont {J.}~\bibnamefont {Chen}},\
  }\bibfield  {title} {\enquote {\bibinfo {title} {{Quantum backflow in
  solutions to the Dirac equation of the spin-1/2 free particle}},}\ }\href
  {\doibase 10.1142/S0217732318501869} {\bibfield  {journal} {\bibinfo
  {journal} {Mod. Phys. Lett. A}\ }\textbf {\bibinfo {volume} {33}},\ \bibinfo
  {pages} {1850186} (\bibinfo {year} {2018})}\BibitemShut {NoStop}%
\bibitem [{\citenamefont {Ashfaque}\ \emph {et~al.}(2019)\citenamefont
  {Ashfaque}, \citenamefont {Lynch},\ and\ \citenamefont
  {Strange}}]{ALS19Relativistic}%
  \BibitemOpen
  \bibfield  {author} {\bibinfo {author} {\bibfnamefont {J.~M.}\ \bibnamefont
  {Ashfaque}}, \bibinfo {author} {\bibfnamefont {J.}~\bibnamefont {Lynch}}, \
  and\ \bibinfo {author} {\bibfnamefont {P.}~\bibnamefont {Strange}},\
  }\bibfield  {title} {\enquote {\bibinfo {title} {{Relativistic quantum
  backflow}},}\ }\href {\doibase 10.1088/1402-4896/ab265c} {\bibfield
  {journal} {\bibinfo  {journal} {Phys. Scr.}\ }\textbf {\bibinfo {volume}
  {94}},\ \bibinfo {pages} {125107} (\bibinfo {year} {2019})}\BibitemShut
  {NoStop}%
\bibitem [{\citenamefont {I.~Bialynicki-Birula}\ and\ \citenamefont
  {Augustynowicz}(2022)}]{BBA22Backflow}%
  \BibitemOpen
  \bibfield  {author} {\bibinfo {author} {\bibfnamefont {Z.~Bialynicka-Birula}\
  \bibnamefont {I.~Bialynicki-Birula}}\ and\ \bibinfo {author} {\bibfnamefont
  {S.}~\bibnamefont {Augustynowicz}},\ }\bibfield  {title} {\enquote {\bibinfo
  {title} {{Backflow in relativistic wave equations}},}\ }\href {\doibase
  10.1088/1751-8121/ac65c1} {\bibfield  {journal} {\bibinfo  {journal} {J.
  Phys. A: Math. Theor.}\ }\textbf {\bibinfo {volume} {55}},\ \bibinfo {pages}
  {255702} (\bibinfo {year} {2022})}\BibitemShut {NoStop}%
\bibitem [{\citenamefont {{Di Bari}}\ \emph {et~al.}(2023)\citenamefont {{Di
  Bari}}, \citenamefont {Paccoia}, \citenamefont {Panella},\ and\ \citenamefont
  {Roy}}]{BPPR23Quantum}%
  \BibitemOpen
  \bibfield  {author} {\bibinfo {author} {\bibfnamefont {L.}~\bibnamefont {{Di
  Bari}}}, \bibinfo {author} {\bibfnamefont {V.~D.}\ \bibnamefont {Paccoia}},
  \bibinfo {author} {\bibfnamefont {O.}~\bibnamefont {Panella}}, \ and\
  \bibinfo {author} {\bibfnamefont {P.}~\bibnamefont {Roy}},\ }\bibfield
  {title} {\enquote {\bibinfo {title} {{Quantum backflow for a massless Dirac
  fermion on a ring}},}\ }\href {\doibase 10.1016/j.physleta.2023.128831}
  {\bibfield  {journal} {\bibinfo  {journal} {Phys. Lett. A}\ }\textbf
  {\bibinfo {volume} {474}},\ \bibinfo {pages} {128831} (\bibinfo {year}
  {2023})}\BibitemShut {NoStop}%
\bibitem [{\citenamefont {Biswas}\ and\ \citenamefont
  {Ghosh}(2021)}]{BG21Quantum}%
  \BibitemOpen
  \bibfield  {author} {\bibinfo {author} {\bibfnamefont {D.}~\bibnamefont
  {Biswas}}\ and\ \bibinfo {author} {\bibfnamefont {S.}~\bibnamefont {Ghosh}},\
  }\bibfield  {title} {\enquote {\bibinfo {title} {{Quantum backflow across a
  black hole horizon in a toy model approach}},}\ }\href {\doibase
  10.1103/PhysRevD.104.104061} {\bibfield  {journal} {\bibinfo  {journal}
  {Phys. Rev. D}\ }\textbf {\bibinfo {volume} {104}},\ \bibinfo {pages}
  {104061} (\bibinfo {year} {2021})}\BibitemShut {NoStop}%
\bibitem [{\citenamefont {Yearsley}\ \emph {et~al.}(2012)\citenamefont
  {Yearsley}, \citenamefont {Halliwell}, \citenamefont {Hartshorn},\ and\
  \citenamefont {Whitby}}]{YHHW12Analytical}%
  \BibitemOpen
  \bibfield  {author} {\bibinfo {author} {\bibfnamefont {J.~M.}\ \bibnamefont
  {Yearsley}}, \bibinfo {author} {\bibfnamefont {J.~J.}\ \bibnamefont
  {Halliwell}}, \bibinfo {author} {\bibfnamefont {R.}~\bibnamefont
  {Hartshorn}}, \ and\ \bibinfo {author} {\bibfnamefont {A.}~\bibnamefont
  {Whitby}},\ }\bibfield  {title} {\enquote {\bibinfo {title} {{Analytical
  examples, measurement models, and classical limit of quantum backflow}},}\
  }\href {\doibase 10.1103/PhysRevA.86.042116} {\bibfield  {journal} {\bibinfo
  {journal} {Phys. Rev. A}\ }\textbf {\bibinfo {volume} {86}},\ \bibinfo
  {pages} {042116} (\bibinfo {year} {2012})}\BibitemShut {NoStop}%
\bibitem [{\citenamefont {Bracken}\ and\ \citenamefont
  {{McGuire}}()}]{BM--Remarks}%
  \BibitemOpen
  \bibfield  {author} {\bibinfo {author} {\bibfnamefont {A.~J.}\ \bibnamefont
  {Bracken}}\ and\ \bibinfo {author} {\bibfnamefont {J.~B.}\ \bibnamefont
  {{McGuire}}},\ }\href@noop {} {\enquote {\bibinfo {title} {{Remarks on
  Quantum Probability Backflow}},}\ }\bibinfo {note}
  {ArXiv:1608.07644}\BibitemShut {NoStop}%
\bibitem [{\citenamefont {Bracken}(2021)}]{Bra21Probability}%
  \BibitemOpen
  \bibfield  {author} {\bibinfo {author} {\bibfnamefont {A.~J.}\ \bibnamefont
  {Bracken}},\ }\bibfield  {title} {\enquote {\bibinfo {title} {{Probability
  flow for a free particle: new quantum effects}},}\ }\href {\doibase
  10.1088/1402-4896/abdd54} {\bibfield  {journal} {\bibinfo  {journal} {Phys.
  Scr.}\ }\textbf {\bibinfo {volume} {96}},\ \bibinfo {pages} {045201}
  (\bibinfo {year} {2021})}\BibitemShut {NoStop}%
\bibitem [{\citenamefont {Mousavi}\ and\ \citenamefont
  {Miret-Art{\'{e}}s}(2022)}]{MM22Different}%
  \BibitemOpen
  \bibfield  {author} {\bibinfo {author} {\bibfnamefont {S.~V.}\ \bibnamefont
  {Mousavi}}\ and\ \bibinfo {author} {\bibfnamefont {S.}~\bibnamefont
  {Miret-Art{\'{e}}s}},\ }\bibfield  {title} {\enquote {\bibinfo {title}
  {{Different routes to the classical limit of backflow}},}\ }\href {\doibase
  10.1088/1751-8121/aca36e} {\bibfield  {journal} {\bibinfo  {journal} {J.
  Phys. A: Math. Theor.}\ }\textbf {\bibinfo {volume} {55}},\ \bibinfo {pages}
  {475302} (\bibinfo {year} {2022})}\BibitemShut {NoStop}%
\bibitem [{\citenamefont {Halliwell}\ \emph {et~al.}(2013)\citenamefont
  {Halliwell}, \citenamefont {Gillman}, \citenamefont {Lennon}, \citenamefont
  {Patel},\ and\ \citenamefont {Ramirez}}]{HGL+13Quantum}%
  \BibitemOpen
  \bibfield  {author} {\bibinfo {author} {\bibfnamefont {J.~J.}\ \bibnamefont
  {Halliwell}}, \bibinfo {author} {\bibfnamefont {E.}~\bibnamefont {Gillman}},
  \bibinfo {author} {\bibfnamefont {O.}~\bibnamefont {Lennon}}, \bibinfo
  {author} {\bibfnamefont {M.}~\bibnamefont {Patel}}, \ and\ \bibinfo {author}
  {\bibfnamefont {I.}~\bibnamefont {Ramirez}},\ }\bibfield  {title} {\enquote
  {\bibinfo {title} {{Quantum backflow states from eigenstates of the
  regularized current operator}},}\ }\href {\doibase
  10.1088/1751-8113/46/47/475303} {\bibfield  {journal} {\bibinfo  {journal}
  {J. Phys. A: Math. Theor.}\ }\textbf {\bibinfo {volume} {46}},\ \bibinfo
  {pages} {475303} (\bibinfo {year} {2013})}\BibitemShut {NoStop}%
\bibitem [{\citenamefont {Chremmos}(2024)}]{Chr24Design}%
  \BibitemOpen
  \bibfield  {author} {\bibinfo {author} {\bibfnamefont {I.}~\bibnamefont
  {Chremmos}},\ }\bibfield  {title} {\enquote {\bibinfo {title} {{Design of
  quantum backflow in the complex plane}},}\ }\href {\doibase
  10.1088/1751-8121/ad1aca} {\bibfield  {journal} {\bibinfo  {journal} {J.
  Phys. A: Math. Theor.}\ }\textbf {\bibinfo {volume} {57}},\ \bibinfo {pages}
  {055301} (\bibinfo {year} {2024})}\BibitemShut {NoStop}%
\bibitem [{\citenamefont {Miller}\ \emph {et~al.}(2021)\citenamefont {Miller},
  \citenamefont {Yuan}, \citenamefont {Dumke},\ and\ \citenamefont
  {Paterek}}]{MYDP21Experiment}%
  \BibitemOpen
  \bibfield  {author} {\bibinfo {author} {\bibfnamefont {M.}~\bibnamefont
  {Miller}}, \bibinfo {author} {\bibfnamefont {W.~C.}\ \bibnamefont {Yuan}},
  \bibinfo {author} {\bibfnamefont {R.}~\bibnamefont {Dumke}}, \ and\ \bibinfo
  {author} {\bibfnamefont {T.}~\bibnamefont {Paterek}},\ }\bibfield  {title}
  {\enquote {\bibinfo {title} {{Experiment-friendly formulation of quantum
  backflow}},}\ }\href {\doibase 10.22331/q-2021-01-11-379} {\bibfield
  {journal} {\bibinfo  {journal} {Quantum}\ }\textbf {\bibinfo {volume} {5}},\
  \bibinfo {pages} {379} (\bibinfo {year} {2021})}\BibitemShut {NoStop}%
\bibitem [{\citenamefont {Barbier}\ and\ \citenamefont
  {Goussev}(2021)}]{BG21experiment}%
  \BibitemOpen
  \bibfield  {author} {\bibinfo {author} {\bibfnamefont {M.}~\bibnamefont
  {Barbier}}\ and\ \bibinfo {author} {\bibfnamefont {A.}~\bibnamefont
  {Goussev}},\ }\bibfield  {title} {\enquote {\bibinfo {title} {{On the
  experiment-friendly formulation of quantum backflow}},}\ }\href {\doibase
  10.22331/q-2021-09-07-536} {\bibfield  {journal} {\bibinfo  {journal}
  {Quantum}\ }\textbf {\bibinfo {volume} {5}},\ \bibinfo {pages} {536}
  (\bibinfo {year} {2021})}\BibitemShut {NoStop}%
\bibitem [{\citenamefont {Goussev}(2019)}]{Gou19Equivalence}%
  \BibitemOpen
  \bibfield  {author} {\bibinfo {author} {\bibfnamefont {A.}~\bibnamefont
  {Goussev}},\ }\bibfield  {title} {\enquote {\bibinfo {title} {{Equivalence
  between quantum backflow and classically forbidden probability flow in a
  diffraction-in-time problem}},}\ }\href {\doibase 10.1103/PhysRevA.99.043626}
  {\bibfield  {journal} {\bibinfo  {journal} {Phys. Rev. A}\ }\textbf {\bibinfo
  {volume} {99}},\ \bibinfo {pages} {043626} (\bibinfo {year}
  {2019})}\BibitemShut {NoStop}%
\bibitem [{\citenamefont {{van Dijk}}\ and\ \citenamefont
  {Toyama}(2019)}]{DT19Decay}%
  \BibitemOpen
  \bibfield  {author} {\bibinfo {author} {\bibfnamefont {W.}~\bibnamefont {{van
  Dijk}}}\ and\ \bibinfo {author} {\bibfnamefont {F.~M.}\ \bibnamefont
  {Toyama}},\ }\bibfield  {title} {\enquote {\bibinfo {title} {{Decay of a
  quasistable quantum system and quantum backflow}},}\ }\href {\doibase
  10.1103/PhysRevA.100.052101} {\bibfield  {journal} {\bibinfo  {journal}
  {Phys. Rev. A}\ }\textbf {\bibinfo {volume} {100}},\ \bibinfo {pages}
  {052101} (\bibinfo {year} {2019})}\BibitemShut {NoStop}%
\bibitem [{\citenamefont {Goussev}(2020)}]{Gou20Probability}%
  \BibitemOpen
  \bibfield  {author} {\bibinfo {author} {\bibfnamefont {A.}~\bibnamefont
  {Goussev}},\ }\bibfield  {title} {\enquote {\bibinfo {title} {{Probability
  backflow for correlated quantum states}},}\ }\href {\doibase
  10.1103/PhysRevResearch.2.033206} {\bibfield  {journal} {\bibinfo  {journal}
  {Phys. Rev. Research}\ }\textbf {\bibinfo {volume} {2}},\ \bibinfo {pages}
  {033206} (\bibinfo {year} {2020})}\BibitemShut {NoStop}%
\bibitem [{\citenamefont {Strange}(2024)}]{Str24Quantum}%
  \BibitemOpen
  \bibfield  {author} {\bibinfo {author} {\bibfnamefont {P.}~\bibnamefont
  {Strange}},\ }\bibfield  {title} {\enquote {\bibinfo {title} {{Quantum
  backflow for a free-particle hermite wavepacket}},}\ }\href {\doibase
  10.1088/1402-4896/ad1ada} {\bibfield  {journal} {\bibinfo  {journal} {Phys.
  Scr.}\ }\textbf {\bibinfo {volume} {99}},\ \bibinfo {pages} {025017}
  (\bibinfo {year} {2024})}\BibitemShut {NoStop}%
\bibitem [{\citenamefont {Yearsley}\ and\ \citenamefont
  {Halliwell}(2013)}]{YH13introduction}%
  \BibitemOpen
  \bibfield  {author} {\bibinfo {author} {\bibfnamefont {J.~M.}\ \bibnamefont
  {Yearsley}}\ and\ \bibinfo {author} {\bibfnamefont {J.~J.}\ \bibnamefont
  {Halliwell}},\ }\bibfield  {title} {\enquote {\bibinfo {title} {{An
  introduction to the quantum backflow effect}},}\ }\href {\doibase
  10.1088/1742-6596/442/1/012055} {\bibfield  {journal} {\bibinfo  {journal}
  {J. Phys. Conf. Ser.}\ }\textbf {\bibinfo {volume} {442}},\ \bibinfo {pages}
  {012055} (\bibinfo {year} {2013})}\BibitemShut {NoStop}%
\bibitem [{\citenamefont {Bracken}\ and\ \citenamefont
  {Melloy}(2014)}]{BM14Waiting}%
  \BibitemOpen
  \bibfield  {author} {\bibinfo {author} {\bibfnamefont {A.~J.}\ \bibnamefont
  {Bracken}}\ and\ \bibinfo {author} {\bibfnamefont {G.~F.}\ \bibnamefont
  {Melloy}},\ }\bibfield  {title} {\enquote {\bibinfo {title} {{Waiting for the
  quantum bus: The flow of negative probability}},}\ }\href {\doibase
  10.1016/j.shpsb.2014.09.001} {\bibfield  {journal} {\bibinfo  {journal}
  {Stud. Hist. Philos. Sci. Part B Stud. Hist. Philos. Mod. Phys.}\ }\textbf
  {\bibinfo {volume} {48}},\ \bibinfo {pages} {13} (\bibinfo {year}
  {2014})}\BibitemShut {NoStop}%
\bibitem [{\citenamefont {Barbier}\ \emph
  {et~al.}(2023{\natexlab{b}})\citenamefont {Barbier}, \citenamefont {Fewster},
  \citenamefont {Goussev}, \citenamefont {Morozov},\ and\ \citenamefont
  {Srivastava}}]{BFG+23Comment}%
  \BibitemOpen
  \bibfield  {author} {\bibinfo {author} {\bibfnamefont {M.}~\bibnamefont
  {Barbier}}, \bibinfo {author} {\bibfnamefont {C.~J.}\ \bibnamefont
  {Fewster}}, \bibinfo {author} {\bibfnamefont {A.}~\bibnamefont {Goussev}},
  \bibinfo {author} {\bibfnamefont {G.~V.}\ \bibnamefont {Morozov}}, \ and\
  \bibinfo {author} {\bibfnamefont {S.~C.~L.}\ \bibnamefont {Srivastava}},\
  }\bibfield  {title} {\enquote {\bibinfo {title} {{Comment on `Backflow in
  relativistic wave equations'}},}\ }\href {\doibase 10.1088/1751-8121/acba62}
  {\bibfield  {journal} {\bibinfo  {journal} {J. Phys. A: Math. Theor.}\
  }\textbf {\bibinfo {volume} {56}},\ \bibinfo {pages} {138003} (\bibinfo
  {year} {2023}{\natexlab{b}})}\BibitemShut {NoStop}%
\bibitem [{\citenamefont {Bracken}\ and\ \citenamefont
  {Melloy}(2023)}]{BM23Comment}%
  \BibitemOpen
  \bibfield  {author} {\bibinfo {author} {\bibfnamefont {A.~J.}\ \bibnamefont
  {Bracken}}\ and\ \bibinfo {author} {\bibfnamefont {G.~F.}\ \bibnamefont
  {Melloy}},\ }\bibfield  {title} {\enquote {\bibinfo {title} {{Comment on
  `Backflow in relativistic wave equations'}},}\ }\href {\doibase
  10.1088/1751-8121/acbd70} {\bibfield  {journal} {\bibinfo  {journal} {J.
  Phys. A: Math. Theor.}\ }\textbf {\bibinfo {volume} {56}},\ \bibinfo {pages}
  {138002} (\bibinfo {year} {2023})}\BibitemShut {NoStop}%
\bibitem [{\citenamefont {Palmero}\ \emph {et~al.}(2013)\citenamefont
  {Palmero}, \citenamefont {Torrontegui}, \citenamefont {Muga},\ and\
  \citenamefont {Modugno}}]{PTMM13Detecting}%
  \BibitemOpen
  \bibfield  {author} {\bibinfo {author} {\bibfnamefont {M.}~\bibnamefont
  {Palmero}}, \bibinfo {author} {\bibfnamefont {E.}~\bibnamefont
  {Torrontegui}}, \bibinfo {author} {\bibfnamefont {J.~G.}\ \bibnamefont
  {Muga}}, \ and\ \bibinfo {author} {\bibfnamefont {M.}~\bibnamefont
  {Modugno}},\ }\bibfield  {title} {\enquote {\bibinfo {title} {{Detecting
  quantum backflow by the density of a Bose-Einstein condensate}},}\ }\href
  {\doibase 10.1103/PhysRevA.87.053618} {\bibfield  {journal} {\bibinfo
  {journal} {Phys. Rev. A}\ }\textbf {\bibinfo {volume} {87}},\ \bibinfo
  {pages} {053618} (\bibinfo {year} {2013})}\BibitemShut {NoStop}%
\bibitem [{\citenamefont {Mardonov}\ \emph {et~al.}(2014)\citenamefont
  {Mardonov}, \citenamefont {Palmero}, \citenamefont {Modugno}, \citenamefont
  {Sherman},\ and\ \citenamefont {Muga}}]{MPM+14Interference}%
  \BibitemOpen
  \bibfield  {author} {\bibinfo {author} {\bibfnamefont {S.}~\bibnamefont
  {Mardonov}}, \bibinfo {author} {\bibfnamefont {M.}~\bibnamefont {Palmero}},
  \bibinfo {author} {\bibfnamefont {M.}~\bibnamefont {Modugno}}, \bibinfo
  {author} {\bibfnamefont {E.~Y.}\ \bibnamefont {Sherman}}, \ and\ \bibinfo
  {author} {\bibfnamefont {J.~G.}\ \bibnamefont {Muga}},\ }\bibfield  {title}
  {\enquote {\bibinfo {title} {{Interference of spin-orbit–coupled
  Bose-Einstein condensates}},}\ }\href {\doibase 10.1209/0295-5075/106/60004}
  {\bibfield  {journal} {\bibinfo  {journal} {EPL (Europhys. Lett.)}\ }\textbf
  {\bibinfo {volume} {106}},\ \bibinfo {pages} {60004} (\bibinfo {year}
  {2014})}\BibitemShut {NoStop}%
\bibitem [{\citenamefont {Eliezer}\ \emph {et~al.}(2020)\citenamefont
  {Eliezer}, \citenamefont {Zacharias},\ and\ \citenamefont
  {Bahabad}}]{EZB20Observation}%
  \BibitemOpen
  \bibfield  {author} {\bibinfo {author} {\bibfnamefont {Y.}~\bibnamefont
  {Eliezer}}, \bibinfo {author} {\bibfnamefont {T.}~\bibnamefont {Zacharias}},
  \ and\ \bibinfo {author} {\bibfnamefont {A.}~\bibnamefont {Bahabad}},\
  }\bibfield  {title} {\enquote {\bibinfo {title} {{Observation of optical
  backflow}},}\ }\href {\doibase 10.1364/OPTICA.371494} {\bibfield  {journal}
  {\bibinfo  {journal} {Optica}\ }\textbf {\bibinfo {volume} {7}},\ \bibinfo
  {pages} {72} (\bibinfo {year} {2020})}\BibitemShut {NoStop}%
\bibitem [{\citenamefont {Daniel}\ \emph {et~al.}(2022)\citenamefont {Daniel},
  \citenamefont {Ghosh}, \citenamefont {Gorzkowski},\ and\ \citenamefont
  {Lapkiewicz}}]{DGGL22Demonstrating}%
  \BibitemOpen
  \bibfield  {author} {\bibinfo {author} {\bibfnamefont {A.}~\bibnamefont
  {Daniel}}, \bibinfo {author} {\bibfnamefont {B.}~\bibnamefont {Ghosh}},
  \bibinfo {author} {\bibfnamefont {B.}~\bibnamefont {Gorzkowski}}, \ and\
  \bibinfo {author} {\bibfnamefont {R.}~\bibnamefont {Lapkiewicz}},\ }\bibfield
   {title} {\enquote {\bibinfo {title} {{Demonstrating backflow in classical
  two beams' interference}},}\ }\href {\doibase 10.1088/1367-2630/aca70b}
  {\bibfield  {journal} {\bibinfo  {journal} {New J. Phys.}\ }\textbf {\bibinfo
  {volume} {24}},\ \bibinfo {pages} {123011} (\bibinfo {year}
  {2022})}\BibitemShut {NoStop}%
\bibitem [{\citenamefont {Ghosh}\ \emph {et~al.}(2023)\citenamefont {Ghosh},
  \citenamefont {Daniel}, \citenamefont {Gorzkowski},\ and\ \citenamefont
  {Lapkiewicz}}]{GDGL23Azimuthal}%
  \BibitemOpen
  \bibfield  {author} {\bibinfo {author} {\bibfnamefont {B.}~\bibnamefont
  {Ghosh}}, \bibinfo {author} {\bibfnamefont {A.}~\bibnamefont {Daniel}},
  \bibinfo {author} {\bibfnamefont {B.}~\bibnamefont {Gorzkowski}}, \ and\
  \bibinfo {author} {\bibfnamefont {R.}~\bibnamefont {Lapkiewicz}},\ }\bibfield
   {title} {\enquote {\bibinfo {title} {{Azimuthal backflow in light carrying
  orbital angular momentum}},}\ }\href {\doibase 10.1364/OPTICA.495710}
  {\bibfield  {journal} {\bibinfo  {journal} {Optica}\ }\textbf {\bibinfo
  {volume} {10}},\ \bibinfo {pages} {1217} (\bibinfo {year}
  {2023})}\BibitemShut {NoStop}%
\bibitem [{\citenamefont {Goussev}\ and\ \citenamefont
  {Joo}(2024)}]{GJ24Simulating}%
  \BibitemOpen
  \bibfield  {author} {\bibinfo {author} {\bibfnamefont {A.}~\bibnamefont
  {Goussev}}\ and\ \bibinfo {author} {\bibfnamefont {J.}~\bibnamefont {Joo}},\
  }\bibfield  {title} {\enquote {\bibinfo {title} {{Simulating quantum backflow
  on a quantum computer}},}\ }\href {\doibase 10.1088/1402-4896/ad2be7}
  {\bibfield  {journal} {\bibinfo  {journal} {Phys. Scr.}\ }\textbf {\bibinfo
  {volume} {99}},\ \bibinfo {pages} {045104} (\bibinfo {year}
  {2024})}\BibitemShut {NoStop}%
\bibitem [{Sup()}]{SupplementalMaterial}%
  \BibitemOpen
  \href@noop {} {}\bibinfo {note} {{See Supplemental Material at \url{https://journals.aps.org/pra/supplemental/10.1103/PhysRevA.110.022216} for the expansion coefficients of the
  backflow-maximizing state.}}\BibitemShut {Stop}%
\bibitem [{\citenamefont {Higuchi}(1988)}]{Hig88Approach}%
  \BibitemOpen
  \bibfield  {author} {\bibinfo {author} {\bibfnamefont {T.}~\bibnamefont
  {Higuchi}},\ }\bibfield  {title} {\enquote {\bibinfo {title} {{Approach to an
  irregular time series on the basis of the fractal theory}},}\ }\href
  {\doibase 10.1016/0167-2789(88)90081-4} {\bibfield  {journal} {\bibinfo
  {journal} {Physica D}\ }\textbf {\bibinfo {volume} {31}},\ \bibinfo {pages}
  {277} (\bibinfo {year} {1988})}\BibitemShut {NoStop}%
\bibitem [{\citenamefont {Liehr}\ and\ \citenamefont
  {Massopust}(2020)}]{LM20On}%
  \BibitemOpen
  \bibfield  {author} {\bibinfo {author} {\bibfnamefont {L.}~\bibnamefont
  {Liehr}}\ and\ \bibinfo {author} {\bibfnamefont {P.}~\bibnamefont
  {Massopust}},\ }\bibfield  {title} {\enquote {\bibinfo {title} {{On the
  mathematical validity of the Higuchi method}},}\ }\href {\doibase
  10.1016/j.physd.2019.132265} {\bibfield  {journal} {\bibinfo  {journal}
  {Physica D}\ }\textbf {\bibinfo {volume} {402}},\ \bibinfo {pages} {132265}
  (\bibinfo {year} {2020})}\BibitemShut {NoStop}%
\bibitem [{\citenamefont {Berry}(1996)}]{Ber96Quantum}%
  \BibitemOpen
  \bibfield  {author} {\bibinfo {author} {\bibfnamefont {M.~V.}\ \bibnamefont
  {Berry}},\ }\bibfield  {title} {\enquote {\bibinfo {title} {{Quantum fractals
  in boxes}},}\ }\href {\doibase 10.1088/0305-4470/29/20/016} {\bibfield
  {journal} {\bibinfo  {journal} {J. Phys. A: Math. Gen.}\ }\textbf {\bibinfo
  {volume} {29}},\ \bibinfo {pages} {6617} (\bibinfo {year}
  {1996})}\BibitemShut {NoStop}%
\end{thebibliography}
\end{document}